\documentclass{jfm}
\usepackage{amsmath, amsfonts, amssymb, bm}
\usepackage{siunitx}
\usepackage{graphicx}
\usepackage{tikz}
\usepackage{xcolor}
\usepackage{epstopdf, epsfig}
\usepackage[acronym]{glossaries}
\usepackage{comment}
\usepackage{longtable}
\usepackage{threeparttablex}
\usepackage[colorlinks=true,
            linkcolor=blue,
            citecolor=blue,
            urlcolor=blue]{hyperref}

\shorttitle{Generalised least squares approach for estimation of the log-law parameters}
\shortauthor{M. Aguiar Ferreira and B. Ganapathisubramani}

\title{Generalised least squares approach for estimation of the log-law parameters of turbulent boundary layers}

\author{M. Aguiar Ferreira\aff{1}
    \corresp{\email{M.Aguiar-Ferreira@soton.ac.uk}},
    \and B. Ganapathisubramani\aff{1}}

\affiliation{\aff{1}Department of Aeronautics and Astronautics, University of
Southampton, Southampton, UK}

\begin{document}

\maketitle

\begin{abstract}
Uncertainty in estimating the log-law parameters is arguably the greatest obstacle to establishing definitive conclusions regarding their numerical values and universality.
This challenge is exacerbated by the limited number of studies that provide thorough uncertainty analyses of experimental data and fitting procedures, and those that do often adopt different approaches, undermining direct comparisons.
The present study applies the generalised least squares (GLS) principle to the log-law velocity profile to establish a standardised, comprehensive framework for quantifying uncertainty in the 
log-law parameters across datasets.
GLS contrasts with ordinary least squares (OLS) and weighted least squares (WLS), which do not account for correlation in errors across measured quantities, as well as with alternative heuristic methods that independently sample primitive variables.
Instead, it incorporates a full covariance matrix of the residuals, propagated from the uncertainties in the primitive variables and consistent with the experimental methods employed.
The study presents a systematic analysis of the response of the log-law regression model using synthetic data, emulating measurements from a hot-wire anemometer mounted on a linear traverse.
This analysis serves as a predictive tool for experimental design, identifying a priori the dominant sources of uncertainty in the log-law parameters and potential mitigation strategies.
The study also provides new insights into the correlation between the log-law parameters and proposes a new fitting procedure that eliminates the need to prescribe the location and extent of the log region.
The open-source Python implementation of the log-law regression model is available for download on GitHub at \url{https://github.com/ma2ferreira/gls_loglaw.git}.
\end{abstract}

\newacronym{cta}{CTA}{constant temperature anemometer}
\newacronym{fe}{FE}{floating element}
\newacronym{gls}{GLS}{generalised least squares}
\newacronym{lda}{LDA}{laser Doppler anemometry}
\newacronym{mle}{MLE}{maximum likelyhood estimation}
\newacronym{ndf}{NDF}{National Diagnostic Facility}
\newacronym{nstap}{NSTAP}{nanoscale thermal anemometry probes}
\newacronym{ntp}{NTP}{normal temperature and pressure}
\newacronym{sse}{SSE}{sum of the squares error}
\newacronym{ofi}{OFI}{oil film interferometry}
\newacronym{ols}{OLS}{ordinary least squares}
\newacronym{piv}{PIV}{particle image velocimetry}
\newacronym{tbl}{TBL}{turbulent boundary layer}
\newacronym{wls}{WLS}{weigthed least squares}
\newacronym{zpg}{ZPG}{zero-pressure-gradient}

\section{Introduction}

The vertical profile of the mean streamwise velocity within the overlap region or inertial sublayer of wall-bounded turbulent flows is often described by a logarithmic distribution, referred to as the \emph{log law} \citep{Prandtl1925}.
It is generally expressed as
\begin{equation}
    \label{eq: log-law}
    U^+ \ = \ \frac{1}{\kappa} \, \log ( z^+) \, + \, A,
\end{equation}
where $U^+=U/U_\tau$ is the inner-normalised mean streamwise velocity, $U_\tau$ is the friction velocity, $z^+=zU_\tau/\nu$ is the inner-normalised wall-normal coordinate, $\nu$ is the kinematic viscosity of the fluid, $\kappa$ is the von Kármán constant, and $A$ is a parameter that depends on the surface condition and is traditionally assumed to be constant for smooth walls.
Estimating the log-law parameters $\kappa$ and $A$ typically involves curve-fitting the data points within the inertial sublayer, a deceptively simple procedure that poses significant challenges.

From a practical standpoint, the inertial sublayer is only loosely defined in classical asymptotic theory.
Accordingly, it emerges in the limit when the inner-outer scale separation, expressed as a ratio by the friction Reynolds number $Re_{\tau} = L / (\nu/U_\tau)$, becomes infinitely large, in the region where the conditions $zU_\tau/\nu \gg 1$ and $z/L \ll 1$ are simultaneously met \citep{Townsend1956}.
The outer scale $L$ depends on the boundary conditions.
It is equivalent to the boundary-layer thickness $\delta$ for boundary-layer flows, the half channel height $H/2$ for channel flows, and the pipe radius $R$ for pipe flows.
This asymptotic definition does not provide a set of objective, quantifiable criteria to unambiguously establish the location of the lower and upper bounds of the inertial sublayer or log region.
Therefore, the choice of suitable empirical criteria becomes an essential part of the curve-fitting procedure, typically left to individual judgement.
This ambiguity represents a source of uncertainty that inevitably propagates to the estimate of the parameters $\kappa$ and $A$.
As evidenced by \citet{Orlu2010}, numerous empirical criteria based on the vertical profile of the mean streamwise velocity have been proposed since \citet{Coles1954}, revealing a stark lack of consensus.

Another difficulty arises from the limited extent of the inertial sublayer up to moderate Reynolds numbers, $Re_{\tau} \lesssim \mathcal{O}(10^4)$, and the spatial resolution of the measurements, which severely limit the number of data points within it and thereby the reduction of statistical uncertainty.
Experimental data of high-Reynolds-number boundary layer, pipe, and channel flows \citep{Hutchins2012, Winkel2012, Hultmark2012, Samie2018} indicate that $Re_{\tau}$ must be well beyond $\mathcal{O}(10^5)$ for the inertial sublayer to span over a decade.
Yet, even in these extreme conditions, the number of data points is seldom over $\sim 10$ and may be considerably lower at the scale of conventional experiments and numerical simulations.

Measurement uncertainty further contributes to the variability in reported values of the log-law parameters.
Its quantification is crucial to establish confidence margins and, most importantly, a basis for comparison between studies.
Although published data are generally complete with the associated uncertainty estimates, different propagation approaches and nuanced interpretations of the diverse sources of error lead to more or less conservative estimates.
This discrepancy can undermine analyses of data sets obtained from multiple sources, such as those of \citet{Zanoun2003}, \citet{Nagib2008}, \citet{Marusic2013}, \citet{Bailey2014}, and \citet{Monkewitz2023}, amongst others, concerning the universality of the log law.
Studies quantifying the uncertainty of traditional measurement methods, including \citet{Thibault2017} and \citet{Rezaeiravesh2018}, play a crucial role in reconciling this discrepancy by providing reference frameworks.
However, failing to make tools available to the wider research community, they fall short of the underlying goal.

As demonstrated by \citet{Segalini2013} and \citet{Bailey2014}, the inference method can also affect the values and the corresponding uncertainty estimates of the log-law parameters. 
\citet{Segalini2013} compared three approaches, namely, direct regression of the log law, the indicator function \citep{Osterlund2000}, and the $\kappa\text{-}A$ scatter method \citep{Alfredsson2013}. 
While they observed marginal discrepancies in the value of the parameters, the uncertainty estimates varied widely.
In particular, the indicator-function approach, which involves differentiation, was found to perform the worst.
\citet{Bailey2014}, focused on pipe-flow measurements and reported estimates obtained from regressions of the log law, the centreline velocity relation, and the friction-factor (bulk) relation, yielding markedly different parameter values and uncertainty estimates. 

Finally, most, and possibly all, studies evaluating the log-law parameters neglect the intrinsic correlation between the primitive (or input) variables, regardless of the fitting method.
Analyses of varying complexity have been conducted, including heuristic approaches such as Monte Carlo simulations and Bayesian hierarchical modelling, yet the primitive variables are invariably sampled independently.
This modelling aspect is particularly important, as the viscous-normalised velocity and wall-normal coordinate are not only autocorrelated but also cross-correlated, both scaled by the friction velocity.
Ignoring these correlations can lead to either over- or underestimation of uncertainty margins, depending on the specific regression structure.
For a comprehensive analysis, it is essential to incorporate the covariance between primitive variables through multivariate sampling or advanced statistical methods.

The present study addresses the challenges mentioned above, applying the \gls{gls} principle introduced by \citet{Aitken1935} and \citet{Sprent1966} to the log-law velocity profile.
\gls{gls} departs from the standard assumptions underpinning the \gls{ols} in which measurement errors in the dependent variable (i.e., $U^+$) are uncorrelated and homoscedastic, i.e. uniformly distributed.
It also differs from the \gls{wls}, where these measurement errors are instead assumed to be uncorrelated but heteroscedastic.
Neither \gls{ols} nor \gls{wls}, the preferred log-law fitting methods, account for the correlation between observational errors.
Additionally, in contrast to heuristic approaches, \gls{gls} is based on an analytical interpretation of uncertainty propagation through the regression model, providing exact solutions to the uncertainty in the log-law parameters and their correlation rather than approximations.
Thus, it provides the opportunity to systematically decouple and isolate the effect of measurement errors in the primitive variables and quantify the influence of flow-dependent factors.

The study aims at providing a standardised and comprehensive framework for uncertainty propagation, and does not attempt to reanalyse published data of boundary-layers, channel, or pipe flows.
As such, the analysis is based on synthetic data, which allows complete control over the underlying parameters.
Following section \ref{sec: model}, where the \gls{gls} regression model is formally defined, section \ref{sec: synthetic} describes the generation of synthetic data, and assesses the response of the model to systematic variations in the uncertainties of the primitive variables, changes in the inner and outer flow scales, the number of degrees of freedom, and the empirical criteria used to define the location and extent of the log region.
Section \ref{sec: insights} presents further insights into current interpretations of the correlations between the log-law parameters and introduces a fitting procedure that is independent of arbitrarily prescribed bounds of the log region. 
Final remarks and key outcomes are summarised in section \ref{sec: conclusion}.
\section{Fitting the log-law velocity profile}
\label{sec: model}

The logarithmic law given in equation~\ref{eq: log-law} may be linearised by introducing the transformed variables $x_i := \log(z_i\,U_\tau/\nu)$ and $y_i := U_i / U_\tau$, for each measurement point $i = 1,...,N$, yielding a linear regression model of the form
\begin{equation}
    \begin{aligned}
        \bm{Y} = \bm{X} \, \bm{\beta} + \bm{\varepsilon}, \\
        \text{with} \quad \bm{X} = \begin{bmatrix} 1 & 1 & \cdots & 1 \\ x_1 & x_2 & \cdots & x_N \end{bmatrix}^\top \quad \text{and} \quad \bm{\beta}=\begin{bmatrix} A & 1/\kappa\end{bmatrix}^\top,
    \end{aligned}
    \label{eq: linear_model}
\end{equation}
where $\bm{Y} \in \mathbb{R}^{N \times 1}$ is the response vector, $\bm{X} \in \mathbb{R}^{N \times 2}$ is the Jacobian matrix of the response vector with respect to the regression coefficients, $\bm{\beta} \in \mathbb{R}^{2 \times 1}$ is the vector of regression coefficients, and $\bm{\varepsilon} \in \mathbb{R}^{N \times 1}$ is the vector of random error terms with zero mean.
Assuming the measurement errors are normally distributed, the maximum likelihood estimate (MLE) for \(\bm{\beta}\) is obtained by minimising the chi-squared merit function
\begin{equation}
    \chi^2(\bm{\beta}) = (\bm{Y} - \bm{X} \, \bm{\beta})^\top \, \bm{W} \, (\bm{Y} - \bm{X} \, \bm{\beta}),
    \label{eq: chi2}
\end{equation}
where 
\(\bm{W} \in \mathbb{R}^{N \times N}\) is a weighting matrix given by the inverse of the covariance matrix of the residuals \(\bm{\varepsilon}\)
\begin{equation}
    \bm{W} = \bm{\Sigma_{\varepsilon}}^{-1}.
\end{equation}
The solution to this minimisation problem yields the expression for the estimate of the regression coefficients,
\begin{equation}
    \bm{\beta} = (\bm{X}^\top \bm{W}  \bm{X})^{-1} \, (\bm{X}^\top \bm{W} \bm{Y}),
\end{equation}
with associated covariance
\begin{equation}
    \bm{\Sigma}_{\beta} = (\bm{X}^\top \bm{W} \bm{X})^{-1}.
    \label{eq: S_beta}
\end{equation}

\subsection{Generalised least squares}
\label{subsec: gls}

\gls{gls} assumes that uncertainties associated with the transformed variables $x$ and $y$ are heteroscedastic and potentially correlated owing to shared sources of measurement error.
These uncertainties may exhibit non-negligible autocorrelation along the vertical profile as well as cross-correlation.
Three primary coupling mechanisms can be identified:
First, uncertainties associated with the streamwise velocity $u$ and the wall-normal coordinate $z$ tend to be autocorrelated, depending on the measurement system, as will become evident in section \ref{sec: synthetic}.
Second, measurement errors in $z$ and the kinematic viscosity $\nu$ directly affect $x$, further contributing to autocorrelated uncertainties along the vertical profile.  
Third, as both transformed variables depend on the friction velocity $U_\tau$, errors in $U_\tau$ introduce negatively correlated uncertainties between the two variables.  
Additional secondary coupling mechanisms arise from measurement errors in air properties, such as temperature, relative humidity, and atmospheric pressure, which propagate into the measurements of $U_\tau$ and $\nu$.

To construct the covariance matrix of the residuals $\bm{\Sigma}_{\varepsilon}$ for a bivariate fit, let $\bm{\omega} = \begin{bmatrix} x_1 & \cdots & x_N & y_1 & \cdots & y_N \end{bmatrix} \in \mathbb{R}^{2N}$
be the vector of $n$ transformed observations and $\bm{\Sigma}_\omega \in \mathbb{R}^{2N \times 2N}$ denote the corresponding covariance matrix.  
Propagating the uncertainty in the transformed variables through a first-order Taylor series expansion,
\begin{equation}
    \bm{\Sigma}_{\varepsilon} = \bm{J}_{\varepsilon} \, \bm{\Sigma}_\omega \, \bm{J}_{\varepsilon}^\top,
    \label{eq: S_epsilon}
\end{equation}
where $\bm{J}_{\varepsilon} \in \mathbb{R}^{N \times 2N}$ is the Jacobian of $\bm{\varepsilon}$ respect to $\bm{\omega}$, written in block from as
\begin{equation}
    \bm{J}_{\varepsilon} = \frac{\partial \bm{\varepsilon}}{\partial \bm{\omega}} = \begin{bmatrix}
    -\beta_1 \, \bm{I}_N & \bm{I}_N
\end{bmatrix}.
\label{eq: J_epsilon}
\end{equation}
As evidenced by equation \ref{eq: J_epsilon}, the Jacobian matrix $\bm{J}_{\varepsilon}$ is a function of the fit parameter $\beta_1$, which is unknown a priori.
Without any prior knowledge of the solution, the linear regression would have to be solved iteratively based on an initial estimate of $\beta_1$, updating $\bm{J}_{\varepsilon}$ until convergence is achieved.
However, in practice, setting $\beta_1 = 1/0.384 $, consistent with previously published studies, already yields a sufficiently converged solution with no further iterations.
Small deviations in the initial estimate of $\beta_1$ have a negligible effect on the resulting uncertainty estimates of the regression coefficients.

Computation of the propagated covariance matrix $\bm{\Sigma}_{\varepsilon}$, expressed by equation~\ref{eq: S_epsilon}, depends on the covariance matrix of the transformed observations $\bm{\Sigma}_\omega$.
This matrix is derived by propagating the uncertainties of the underlying primitive variables, which include the wall-normal coordinate, streamwise velocity, friction velocity, and kinematic viscosity.
Accordingly, let $\boldsymbol{\theta} =
\begin{bmatrix} z_1 & \cdots & z_N & U_1 & \cdots & U_N & U_\tau & \nu \end{bmatrix} \in \mathbb{R}^{2N+2}$, be the vector of primitive variables, and $\bm{\Sigma}_\theta \in \mathbb{R}^{(2N+2) \times (2N+2)}$ be the corresponding covariance matrix.
It follows that the covariance matrix of the transformed observations is approximated as
\begin{equation}
\bm{\Sigma}_\omega = \bm{J}_\omega \, \bm{\Sigma}_\theta \, \bm{J}_\omega^\top,
\end{equation}
where $\bm{J}_\omega \in \mathbb{R}^{2N \times (2N+2)}$ is the Jacobian matrix of $\bm{\omega}$ with respect to $\boldsymbol{\theta}$.  
The Jacobian matrix is explicitly expressed in block form as
\begin{equation}
\bm{J}_\omega = \frac{\partial \bm{\omega}}{\partial \boldsymbol{\theta}} = \begin{bmatrix}
        \text{diag}(\bm{z}) & 
        \bm{0}_{N \times N} & 
        \displaystyle \dfrac{1}{U_\tau} \bm{1}_{N} & 
        \displaystyle -\dfrac{1}{\nu} \bm{1}_{N} \\[4mm]
        \bm{0}_{N \times N} & 
        \displaystyle \dfrac{1}{U_\tau} \bm{I}_N & 
        \displaystyle -\dfrac{1}{U_\tau^2} \bm{U} & 
        \bm{0}_{N}
    \end{bmatrix},
\end{equation}
where $\bm{z} = \begin{bmatrix} z_1 & \cdots & z_N \end{bmatrix}^\top$ and $\bm{u} = \begin{bmatrix} U_1 & \cdots & U_N \end{bmatrix}^\top$.
The covariance matrix of the primitive variables is defined as
\begin{equation}
    \bm{\Sigma}_\theta =
    \begin{bmatrix}
        \bm{\Sigma}_{z} & \bm{\Sigma}_{zU} & \bm{\Sigma}_{z U_\tau} & \bm{\Sigma}_{z \nu} \\[1mm]
        \bm{\Sigma}_{Uz} & \bm{\Sigma}_{U} & \bm{\Sigma}_{U U_\tau} & \bm{\Sigma}_{U \nu} \\[1mm]
        \bm{\Sigma}_{U_\tau z} & \bm{\Sigma}_{U_\tau U} & \sigma^2_{U_\tau} & \sigma_{U_\tau \nu} \\[1mm]
        \bm{\Sigma}_{\nu z} & \bm{\Sigma}_{\nu U} & \sigma_{\nu U_\tau} & \sigma^2_{\nu}
        \end{bmatrix},
    \label{eq: S_theta}
\end{equation}
where $\bm\Sigma$ denotes a covariance matrix between vectors of variables, $\sigma^2$ denotes the variance of individual scalar variables, and $\sigma_{U_\tau\nu}$ denotes the covariance between the friction velocity $U_\tau$ and the kinematic viscosity $\nu$.
Estimating $\bm{\Sigma}_\theta$ therefore requires explicit assumptions about the statistical dependence between primitive variables and a careful quantification of the associated uncertainties.
The methodology, described in detail in section~\ref{subsec: baseline}, follows the framework outlined in the guide to the expression of uncertainty in measurement \citep{JCGM2008}.
In this framework, each block of $\bm{\Sigma}_\theta$ comprises statistical, type~A, and systematic, type~B, uncertainty components.
The type~A component represents stochastic variability conditional on the regression model.
It stems from signal fluctuations due to noise, unmonitored variations in environmental conditions over time, or instrument repeatability issues.
The type~B component, in contrast, accounts for structured biases and uncertainties associated with model specification, calibration, and reference standards.

Once $\bm{\Sigma}_\theta$ is specified, the \gls{gls} formulation yields the maximum likelihood estimate of the regression coefficients $\bm\beta$ and the associated covariance $\bm\Sigma_\beta$.
The log-law parameters $\bm{\psi}=[A, \kappa]^\top$ then follow directly from the functional relationship with $\bm\beta$, and the corresponding covariance matrix is obtained by propagating $\bm\Sigma_\beta$ again via a first-order Taylor series expansion,
\begin{equation}
    \bm{\Sigma}_{\psi} = \bm{J}_{\psi} \bm{\Sigma}_\beta \bm{J}_{\psi}^\top,
\end{equation}
where $\bm{J}_{\psi}$ is the Jacobian matrix of $\bm{\psi}$ with respect to $\bm{\beta}$.
The Jacobian matrix is explicitly expressed as
\begin{equation}
    \bm{J}_{\psi} = \frac{\partial\bm{\psi}}{\partial\bm{\beta}} = \begin{bmatrix} 1 & 0 \\ 0 & -1/\beta_1^2\end{bmatrix} = \begin{bmatrix} 1 & 0 \\ 0 & -\kappa^2\end{bmatrix}.
\end{equation}

\subsection{Special cases}
\label{subsec: wls_ols}
Standard variations of the \gls{gls} formulation, including \gls{wls} and \gls{ols}, emerge from modifications to the initial assumptions about the variances and covariances of the residual vector $ \bm{\varepsilon} $ outlined above.
\gls{wls} considers heteroscedastic (unequal variance) but uncorrelated errors in the dependent variable $y_i$, while treating the independent variable $x_i$ as exact.
The residual covariance matrix then reduces to $\bm{\Sigma}_\varepsilon = \text{diag}(\bm{\sigma}_y^{2})$, so that the weighting matrix $\bm{W} = \bm{\Sigma}^{-1}_\varepsilon = \text{diag}(\bm{\sigma}_y^{-2})$ is a function of $\bm\sigma_U^2$ and $\sigma_{U_\tau}^2$.
\gls{ols} similarly treats the independent variable as exact, but further assumes homoscedastic (equal variance) and uncorrelated errors, so that 
$\bm{\Sigma}_\varepsilon = \sigma_y^2 \bm{I}$ with $\sigma_y^2$ unknown.
The weight matrix reduces to $\bm{W} \propto \bm{I}$, and the chi-squared merit function of equation~\ref{eq: chi2} reduces, up to the constant factor $1/\sigma_y^2$, to the \gls{sse} of the fit
\begin{equation}
    SSE = (\bm{Y} - \bm{X}\bm{\beta})^\top (\bm{Y} - \bm{X}\bm{\beta}).
\end{equation}
The corresponding vector of regression coefficients is given by
\begin{equation}
    \bm{\beta} = (\bm{X}^\top \bm{X})^{-1} \, (\bm{X}^\top \bm{Y}),
\end{equation}
with associated covariance
\begin{equation}
    \bm{\Sigma}_\beta = \sigma^2_y \, (\bm{X}^\top \bm{X})^{-1}.
    \label{eq: Sigma_beta_OLS}
\end{equation}
As $\sigma_y^2$ is often unknown, it is common practice to approximate this term by the unbiased sample estimator $\hat{\sigma}_y^2 = SSE/(N-2)$, where $N - 2$ is the number of degrees of freedom for a two-parameter model.

\subsection{\texorpdfstring{$\bm{\Sigma}_\psi$}{} decomposition}
\label{subsec: decomposition}
To examine the influence of the regression model on the statistical (type~A) uncertainty component, this section derives a closed-form decomposition of the covariance matrix of the log-law parameters into geometry-dependent and flow-dependent contributions.
Owing to the structural complexity of the \gls{gls} framework, which involves the coupling of measurement errors across multiple primitive variables, an analytical decomposition of this kind would be nearly intractable. 
Therefore, the \gls{ols} framework is adopted instead to develop intuition and to provide a more detailed interpretation of the results.

Assuming the statistical error associated with the mean velocity estimates is the only source of uncertainty, such that $\sigma_y = \sigma_U/U_\tau$, the \gls{ols} covariance matrix of the fit parameters is explicitly written
\begin{equation}
    \begin{split}
        \bm{\Sigma}_\beta = \frac{\sigma^2_U}{{U^2_\tau\Delta}} \begin{bmatrix} S_2 & -S_1 \\ -S_1 & S_0 \end{bmatrix},  \\ 
        \qquad  {\Delta} = S_0S_2 - S^2_1, \quad S_0 = N, \quad S_1 = \sum_i^N x_i, \quad S_2 = \sum_i^N x^2_i.
    \end{split}
\end{equation}
Propagating to the log-law parameters $\bm\psi = [A, \kappa]$, the associated covariance matrix becomes
\begin{align}
    \bm{\Sigma}_\psi = \frac{\sigma^2_U}{{U^2_\tau\Delta}} \begin{bmatrix} S_2 & \kappa^2 S_1 \\ \kappa^2 S_1 & \kappa^4 S_0 \end{bmatrix}.
\end{align}
Decomposing $x_i$ by substituting $Re_\tau = U_\tau L/\nu$ into its definition,
\begin{equation}
    x_i = \log\left(\frac{z_i} {L}\right) + \log\left(\frac{U_\tau L}{\nu}\right) = \zeta_i + R,
\end{equation}
where $\zeta_i \equiv \log(z_i/L)$ depends only on the outer scaled wall-normal coordinates and $R \equiv \log(Re_\tau)$ is constant across all data points.
Rewriting the sums based on this decomposition yields
\begin{equation}
    \begin{split}
        S_1 =  \sum_i^N \zeta_i + NR = \overline{S}_1 + NR \\
        S_2 =  \sum_i^N x_i^2 = \sum_i^N \zeta^2_i + 2R\sum_i^N \zeta_i + N R^2 = \overline{S}_2 + 2R\overline{S}_1 + NR^2, \\
    \end{split}
\end{equation}
where $S_0=N$ remains unchanged, and $\overline{S}_1$ and $\overline{S}_2$ depend only on the measurement geometry, i.e. the spread or distribution of data points.
Substituting the sums into the determinant and expanding,
\begin{equation}
    \Delta = N(\overline{S}_2 + 2R\overline{S}_1 + NR^2) - (\overline{S}_1 + NR)^2 = N\overline{S}_2 - \overline{S}^2_1.
\end{equation}
Recognising that $N\overline{S}_2 - \overline{S}_1^2 = N\sum_i^N(\zeta_i - \overline{\zeta})^2$, with the mean of $\zeta_i$ given by $\overline{\zeta} = \overline{S}_1/N$, the determinant reduces to
\begin{equation}
    \Delta = N^2\sigma^2_{\zeta}, 
\end{equation}
where the variance $\sigma^2_{\zeta} = (1/N)\sum_i^N(\log(z_i) - \overline{\log(z)})^2$ depends only on the wall-normal coordinates, as $\log(L)$ representing a constant shift cancels itself out.
Substituting the decomposed sums and the determinant into the covariance matrix,
\begin{align}
    \bm{\Sigma}_\psi = \frac{\sigma^2_U}{U^2_\tau N^2\sigma^2_{\zeta}} \begin{bmatrix} \overline{S}_2 + 2R\overline{S}_1 + NR^2 & \kappa^2( \overline{S}_1 + NR) \\ \kappa^2(\overline{S}_1 + NR) & \kappa^4 N \end{bmatrix}.
    \label{eq: Sigma_phi}
\end{align}

For sufficiently large $Re_\tau$, such that $R \gg \overline{\zeta}$, the ratios $\overline{S}_1/R = \mathcal{O}(N\overline{\zeta}/R)$ and $\overline{S}_2/(NR^2) = \mathcal{O}(\overline{\zeta}^2/R^2)$ become asymptotically negligible.
Retaining only the leading-order term in $R$, the covariance matrix reduces to
\begin{align}
    \bm{\Sigma}_\psi \approx \frac{\sigma^2_U}{U^2_\tau N\sigma^2_{\zeta}} \begin{bmatrix} R^2 & \kappa^2 R \\ \kappa^2 R & \kappa^4 \end{bmatrix},
    \label{eq: Sigma_phi_high_Re}
\end{align}
which is a rank--$1$ matrix.
Thus, as $Re_\tau \rightarrow \infty$, $\det(\bm\Sigma_\psi) \rightarrow 0$, and the uncertainties in the log-law parameters become perfectly correlated.
In geometrical terms, this limiting behaviour is equivalent to the log-law profile shifting towards increasingly large $x_i$, such that the data points constrain only a single linear combination of $A$ and $\kappa$.
At the same time, the individual variances and covariance decrease, as becomes evident by recasting the prefactor as $(\sigma^2_U L^2)(Re^2_\tau \nu N\sigma^2_{\zeta})$ and noting that $R^2 = \log^2(Re_\tau)$ grows much more slowly than $Re^2_\tau$.
This effect is further compounded by the consequent increase in the number of observations within the log region $N$ and their spatial distribution $\sigma^2_\zeta$.

Equation~\ref{eq: Sigma_phi_high_Re} highlights a crucial advantage of high-Reynolds-number experiments, which are intrinsically better conditioned for estimating the log-law parameters.
The decrease in uncertainty comes at the cost of a stronger correlation between the estimates, which nonetheless remain identifiable under practical conditions.
While beneficial, this relationship has limited implications as it applies exclusively to statistical sources of uncertainty.
Experiments often involve additional systematic sources of uncertainty that are not accounted for in the \gls{ols} framework and do not scale with this solution, potentially dominating the uncertainty budget under specific conditions.

\section{Application to synthetic data}
\label{sec: synthetic}

This section presents a \emph{qualitative}, sensitive analysis of the uncertainty associated with the log-law parameters based on synthetic data of the velocity profile of turbulent boundary layers.
Although a wealth of high-quality experimental data sets across a broad range of Reynolds numbers is available in the literature \citep[amongst others]{Winkel2012, Hultmark2012, Hutchins2012, Samie2018}, the use of synthetic data provides complete control over the underlying parameters.
This control facilitates a more transparent assessment of the sources of error and propagation of uncertainty, including potential coupling mechanisms between measurement errors.
It is beyond the scope of this study to reanalyse existing data sets or to contribute to the ongoing debate regarding the universality of the log-law parameters.
The framework developed here may nonetheless provide a robust basis for future studies seeking to revisit these issues through a more comprehensive treatment of measurement uncertainty.

\subsection{Generation of synthetic data}
\label{subsec: gen_data}

\begin{table}
    \centering
    \begin{threeparttable}
        \renewcommand{\arraystretch}{1.2}
        \setlength{\tabcolsep}{5mm}
        \begin{tabular}{cccccc}
            $\kappa$ & $a$ & $\Pi$ & $A$\tnote{a} & $\rho$ (\si{\kilogram\per\meter\cubed}) & $\nu$ (\si{\meter\squared\per\second}) \\\hline
            $0.384$ & $-10.3061$ & $0.45$ & $4.17$ & $1.204$ & $1.51\times10^{-5}$
        \end{tabular}
        \begin{tablenotes}
            \item[a] Set implicitly \citep{Monkewitz2007}.
        \end{tablenotes}
    \end{threeparttable}
    \caption{Boundary-layer parameters and air properties, representative of typical experimental conditions, including the von K\'arm\'an constant $\kappa$, the Musker parameter $a$, the additive constant $A$, the air density $\rho$, and kinematic viscosity $\nu$.}
    \label{tab: parameters}
\end{table}

The analysis adopts the established composite formulation for the velocity profile of a \gls{zpg} turbulent boundary layer, which combines the near-wall function of \citet{Musker1979}, the buffer-region bump correction of \citet{Monkewitz2007}, and the Coles-type wake function of \citet{Chauhan2009}.
For brevity, the detailed derivation is omitted.
The formulation expresses $U^+(z^+;\kappa, a, \Pi)$ in terms of the von K\'arm\'an constant $\kappa = 0.384$, the Musker parameter $a = -10.3061$, and the wake parameter $\Pi = 0.45$. 
It yields a smooth, continuous representation of the velocity profile from the edge of the viscous sublayer $z^+ \sim 5$, where $U^+=z^+$, through the inertial sublayer, where $U^+$ follows the log law with an additive constant $A = 4.17$.
As listed in table \ref{tab: parameters}, the boundary-layer parameters $\kappa$, $a$, $A$ (implicitly), $\Pi$, and the air properties $\rho$ and $\nu$, are held constant, while $U_{\tau}$ and $\delta$ are varied systematically to assess the influence of flow scale.

To reproduce the effect of correlated measurement errors in the wall-normal coordinate and the streamwise velocity, synthetic data are generated to emulate measurements from hot-wire anemometry.
This flow diagnostic technique offers unmatched pointwise temporal and spatial resolutions, along with the ability to traverse wide flow regions in arbitrary directions.
For these capabilities, it has become the benchmark experimental method for investigating the turbulent structure and interscale dynamics of wall-bounded flows, and, critically, testing the universality of the log law. 
Following this approach strengthens the impact of the analysis while maintaining a clear focus on the \gls{gls} formulation.
The framework can later be adapted to reproduce correlated errors associated with alternative flow diagnostic techniques, such as \gls{piv} and \gls{lda}.

\begin{figure}
    \centering
    \includegraphics{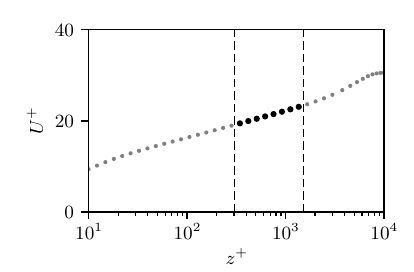}
    \caption{(a) Synthetic hot-wire measurements of the viscous-normalised velocity profile and log-law fit, with parameters listed in table \ref{tab: parameters}, friction Reynolds number $Re_\tau = 10^4$, and boundary-layer thickness $\delta=0.1$\,\si{\meter}.
    The dashed vertical lines indicate the location and extent of the log region $3Re_\tau^{1/2}<z^+<0.15Re_\tau$.}
    \label{fig: profile_baseline}
\end{figure}

Figure \ref{fig: profile_baseline} shows an example of synthetic hot-wire anemometry data for a turbulent boundary layer at a friction Reynolds number $Re_\tau=10^4$.
The vertical profile comprises $30$ logarithmically-spaced points within the range $10 < z^+ < 0.3Re_\tau$ followed by $10$ linearly-spaced points extending to $z^+=\delta^+$.
A total of $8$ points, black-coloured markers, fall within the log region $3Re_\tau^{1/2}<z^+<0.15Re_\tau$ \citep{Marusic2013}, as indicated by the vertical-dashed lines.
This number is characteristic of experimental measurements at typical laboratory scales with $\delta \sim \mathcal{O}(0.1)$\,\si{\meter}.

\subsection{Baseline uncertainty estimates}
\label{subsec: baseline}
Having established a model for the velocity profile and hot-wire anemometry as the basis for this analysis, the covariance of the primitive variables $\bm{\Sigma}_\theta$, given by equation \ref{eq: S_theta}, can now be estimated.
It is subsequently propagated through the regression model to obtain the covariance of the transformed variables $\boldsymbol{\Sigma}_{\theta}$, the covariance of the residuals $\boldsymbol{\Sigma}_{\varepsilon}$, and finally the covariance of the regression coefficients $\boldsymbol{\Sigma}_{\beta}$.
As briefly explained in section \ref{subsec: gls}, the estimation of $\bm{\Sigma}_\theta$ requires either knowledge of, or explicit assumptions about, the statistical dependence of the primitive variables $\bm\theta$ and how their uncertainties are quantified.

The wall-normal coordinate $z$ and the streamwise velocity $U$ are mutually independent, as the vertical traverse and the hot-wire anemometry are two separate systems.
They are also both independent of the friction velocity $U_\tau$ and the kinematic viscosity $\nu$.
In contrast, $U_\tau$ and $\nu$ are mutually correlated because $U_\tau = (\tau_\omega / \rho) ^ {1 / 2}$, where $\tau_\omega$ is the wall-shear stress, and $\nu$ and $\rho$ both depend on the measurement of environmental conditions (temperature, relative humidity, and atmospheric pressure).
A similar coupling mechanism exists between the measurement error of $U$ and that of $U_\tau$ and $\nu$, as calibration of hot-wire anemometry also involves the measurement of environmental conditions.
Neglecting these second-order coupling effects, $\bm{\Sigma}_\theta$ assumes the block-diagonal structure,
\begin{equation}
    \bm{\Sigma}_\theta =
    \begin{bmatrix}
        \bm{\Sigma}_{z} & 0 & 0 & 0 \\[1mm]
        0 & \bm{\Sigma}_{U} & \textcolor{gray}{\bm{\Sigma}_{U U_\tau}} & \textcolor{gray}{\bm{\Sigma}_{U \nu}} \\[1mm]
        0 & \textcolor{gray}{\bm{\Sigma}_{U_\tau U}} & \sigma^2_{U_\tau} & \textcolor{gray}{\sigma_{U_\tau \nu}} \\[1mm]
        0 & \textcolor{gray}{\bm{\Sigma}_{\nu U}} & \textcolor{gray}{\sigma_{\nu U_\tau}} & \sigma^2_{\nu}
    \end{bmatrix},
    \label{eq: S_theta_assumption}
\end{equation}
where neglected terms are shown in grey.
As summarised in table \ref{tab: uncertainty}, baseline uncertainty estimates of type A and type B, representative of typical experimental conditions, are assigned to the primitive variables $\bm\theta$.
In particular, to reproduce the effect of correlated errors, the covariance of the wall-normal coordinate $\bm{\Sigma}_{z}$ and the covariance of the streamwise velocity $\bm{\Sigma}_{U}$ are estimated by explicitly considering the experimental procedures that would be involved in the operation of the vertical traverse and the hot-wire anemometry systems, respectively.
An overview of these procedures and the corresponding modelling assumptions is provided below.

\begin{table}
    \renewcommand{\arraystretch}{1.2}
    \setlength{\tabcolsep}{5mm}
    \centering
    \begin{tabular}{cccp{4.5cm}}
        $\theta$ & $\sigma^A_\theta$ & $\sigma^B_\theta$ & Source (type\,-\,B) \\\hline
        $\bm{z}$ & $0$ & $\bm{\Sigma}_{z}(\sigma_{z_0}, \sigma_{\Delta z})$ & \\
        \qquad $\Delta z$ & $0$ & $1\,\si{\micro\meter}$ & Encoder \\
        \qquad $z_0$ & $0$ & $10\,\si{\micro\meter}$ & Reference wall datum \\
        $\bm{U}$ & $\text{diag}(\bm{\sigma}_U^{2})$ & $\bm{\Sigma}_{U}(\bm{\sigma}_E, \bm{\sigma}_{q_0}, \sigma_\rho)$ & \\
        \qquad $\bm{E}$ & $0$ & $0.001\bm{E}$ & Hot-wire control \\
        \qquad $\bm{q_0}$ & $0$ & $0.002\bm{q_0}$ & Micromanometer \\
        \qquad $\rho$ & $0$ & $0.005\rho$ & \\
        $\nu$ & $0$ & $0.01\nu$ & \\
        $U_\tau$ & $0$ & $0.01 U_\tau$ & OFI or FE balance \\
    \end{tabular}
    \caption{Baseline uncertainties in the primitive variables, representative of values reported in studies that provide a quantitative assessment of the log-law parameters, as discussed in section \ref{subsec: baseline}.
    The contributions of type~A arise from statistical (random) errors and those of type~B from systematic (bias) errors. 
    }
    \label{tab: uncertainty}
\end{table}

\subsubsection{Wall-normal coordinate}
$\bm{\Sigma}_{z}$ depends on the specifications of the hot-wire traverse system and the measurement method of the reference wall datum.
A typical traverse system driven by a stepper motor or a DC motor, free of backlash and combined with an encoder, achieves a relative positional error of approximately $1$\,\si{\micro\meter} per translation $\Delta z$.
This contribution is generally small compared with the uncertainty arising from the measurement of the reference wall datum $z_0$, which can be estimated either by direct contact using datum blocks or through non-contact methods, such as optical or capacitance measurements.
\citet{Hutchins2002} and \citet{Orlu2010} provide comprehensive reviews of these approaches and note that non-contact methods tend to perform better, as they eliminate the risk of probe-holder deformation.
The typical uncertainty level is on the order of $10\,\si{\micro\meter}$, averaged from values reported in the literature.
For example, $5$\,\si{\micro\meter} \citep{Osterlund1999, Zanoun2003, McKeon2004, Zanoun2007, Vallikivi2015}, $20$\,\si{\micro\meter}  \citep[pipe flow,][]{Monty2005}, $50$\,\si{\micro\meter} \citep{Zagarola1998}, and $100$\,\si{\micro\meter} \citep[channel flow,][]{Monty2005}.
In addition to these uncertainty sources, \citet{Hutchins2002} reason that a misalignment between the wall-normal direction and the traverse axis could induce positional drift, skewing the measured velocity profile.
However, this contribution is not factored into the uncertainty propagation analysis, which is detailed in appendix~\ref{app: Sigma_zz}.

\subsubsection{Mean streamwise velocity}
As shown in table~\ref{tab: uncertainty}, the combined uncertainty in the mean streamwise velocity comprises both type~A and type~B components.
It is given by
\begin{equation}
    \bm{\Sigma}_U = \bm{\Sigma}^A_U + \bm{\Sigma}^B_U,
    \label{eq: Sigma_u}
\end{equation}
where $\bm{\Sigma}^A_U = \text{diag}(\bm{\sigma}_U^{2})$ is a diagonal matrix representing the statistical uncertainty in the mean velocity estimates, which depend on the local turbulent flow characteristics and the duration of data acquisition. 
Assuming a representative turbulence intensity in the logarithmic region of $u'_{rms, i}/U_i \sim 0.1$, where $u'_{rms}$ is the root-mean-square of the velocity fluctuations, and a typical acquisition period for hot-wire measurements corresponding to at least $T U_\infty/\delta \sim 10^{4}$ boundary-layer turnover times, the statistical uncertainty in the mean velocity estimates is $\sigma^A_{U,i}/U_i \sim 0.1/(10^4)^{1/2} = 10^{-3}$.
For reference, $0.1$\,\% \citep{Winkel2012}, $0.3$\,\% \citep{Eggels1994, Durst1995, Zagarola1998, McKeon2004}, and $0.6$\,\% for measurements in the atmospheric surface layer \citep{Andreas2006} are typical values.

The second term on the right-hand side of equation \ref{eq: Sigma_u}, representing the systematic uncertainty component, depends on the calibration of the hot-wire probe that relates the flow velocity $U$ to the voltage drop across the wire $E$, using King's law \citep{King1914} or otherwise a higher-order polynomial function.
The in-situ calibration process involves taking a series of measurement pairs of the free-stream dynamic pressure $q_0$ using a Pitot-static probe and the corresponding $E$.
The free-stream velocity is then inferred as $U_0=(2q_0/\rho)^{1/2}$, where $\rho$ is the air density approximated from the ideal gas law given measurements of the air temperature $T$ and the atmospheric pressure $p_{a}$.
Therefore, $q_0$, $E$, $T$, and $p_a$ are the primitive variables whose uncertainties propagate through the calibration model and ultimately determine $\bm{\Sigma}^B_{U}$.

A detailed uncertainty propagation analysis of this calibration process is provided in appendix~\ref{app: Sigma_uu}, introducing four simplifications.
First, King’s law is linearised by prescribing a fixed value for the exponent, representative of \gls{ntp} conditions.
Second, the analysis accounts for the combined uncertainty of the measurement instruments, including both systematic and statistical components.
In particular, \citet{Jorgensen1996} observed that for a standard hot-wire with a nominal resistance of $\sim 1$\,\si{\ohm}, the repeatability error in setting the overheat ratio is approximately $1$\,\si{\milli\ohm}, corresponding to a repeatability error in $E$ of $0.1$\,\%.
A baseline uncertainty of $0.2$\,\% is assigned to $q_0$, consistent with specifications reported by manufacturers of bench-top micromanometers.
Third, the analysis neglects the statistical uncertainty in the reference mean velocity measurements, which is relatively small under controlled calibration conditions, as evidenced by \cite{Rezaeiravesh2018}.
Fourth, $\rho$ is treated as a primitive variable instead of $T$ and $p_a$ and assigned a baseline uncertainty of $1$\,\%.

\subsubsection{Friction velocity}
The uncertainty in the measurement of wall-shear stress, from which the friction velocity can be inferred, is one of the most critical factors in estimating the log-law parameters.
The selection of the measurement method depends on the experimental configuration.
For example, \gls{ofi} has been common practice for boundary-layer and channel flows, yielding uncertainties in the friction velocity around $1.50$\,\% \citep{Osterlund1999} and $1.25$\,\% \citep{Zanoun2003}.
Alternatively, the \gls{fe} balance has been used for boundary-layer flows, achieving uncertainties of $2$\,\% \citep{Winkel2012} and $0.5$\,\% \citep{Baars2016, Ferreira2024} in carefully controlled experiments under nominally \gls{zpg} conditions.
For fully-developed pipe flows, the friction velocity can be inferred from pressure gradient measurements to within an uncertainty of $2$\,\% \citep{Nikuradse1932, Eggels1994} down to $0.5$\,\% \citep{Zagarola1998, McKeon2004, Monty2005}.
The friction velocity has also been estimated from the slope of the velocity profile in the viscous sublayer, $U^+(z^+<5)=z^+$, based on \gls{lda} \citep{Durst1995} and \textmu\gls{piv} \citep{Kahler2005} measurements, yielding uncertainties of approximately $0.5$\,\%.
Preston tube measurements, while popular in early studies, have increasingly fallen out of use, with reported uncertainties of about $1.5$\,\% \citep{Monty2005}.
Studies of atmospheric surface layers typically use indirect estimation methods, for example, based on the Reynolds shear-stress balance, yielding comparably high uncertainties of approximately $5$\,\% \citep{Andreas2006} and $10$\,\% \citep{Hutchins2012}.
Based on these reported values, a relatively optimistic baseline uncertainty of $1$\,\% is assigned to the friction velocity.

\subsubsection{Air properties}
Similarly to air density, the kinematic viscosity $\nu$ depends on $T$ and $p_a$ but is treated as a primitive variable with an equal baseline uncertainty estimate of $1\,$\%.
This simplification implies that any error correlation that would arise from the measurement of environmental conditions is neglected, as indicated in equation \ref{eq: S_theta_assumption}.

\subsection{Sensitivity of model predictions}

\begin{figure}
    \centering
    \includegraphics{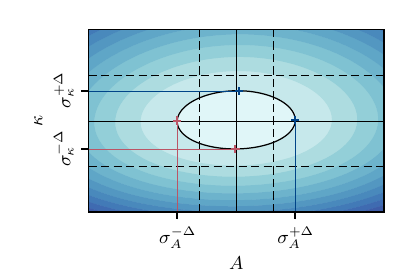}
    \caption{Reduced chi-squared merit function and joint uncertainty region of the log-law fit parameters at a confidence level of $95\%$ for the velocity profile in figure \ref{fig: profile_baseline}.
    The dashed vertical and horizontal lines denote a $\pm5\%$ deviation from the fitted values.
    Uncertainties in the primitive variables correspond to the baseline values listed in table \ref{tab: uncertainty}.}
    \label{fig: u_baseline}
\end{figure}

Figure \ref{fig: u_baseline}(b) shows the joint uncertainty region of the log-law parameters at a confidence level of $95$\,\%, for the baseline uncertainties in the primitive variables listed in table~\ref{tab: uncertainty}.
The dashed vertical and horizontal lines indicate a $5$\,\% deviation from the estimated values $A = 4.094\,(-0.327, +0.327)$ and $\kappa = 0.3815\,(-0.0126,+0.0118)$.
These uncertainty margins are approximately twice and three times larger, respectively, than those obtained solely based on the residuals of the curve fit, as is common practice in various studies listed in table \ref{tab: survey}.
Reported uncertainties in the log-law parameters thus appear to be significantly underestimated.

Expanding upon this solution, the following section examines the response of the \gls{gls} model to systematic variations in the uncertainty of the primitive variables from the baseline values, as well as to changes in the inner and outer flow scales, the number of degrees of freedom, and the empirical criteria used to define the location and extent of the log region.
Because of the multidimensional nature of this parameter space, the analysis does not attempt an exhaustive characterisation of individual or coupled effects, but instead highlights the inherent difficulty of obtaining accurate and robust estimates of log-law parameters under typical experimental conditions.

\subsubsection{Uncertainty in primitive variables}
\label{subsubsec: u_primitives}
\begin{figure}
    \centering
    \includegraphics{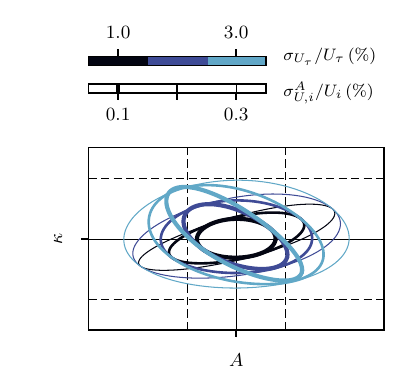}
    \includegraphics{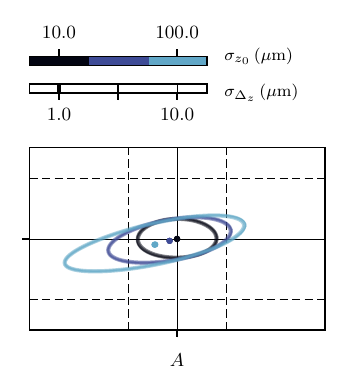}
    \caption{Joint uncertainty regions of the log-law fit parameters at a confidence level of $95\%$, for varying uncertainty in the primitive variables: (a) $U_\tau$ and $U$ and (b) $z_0$ and $\Delta_z$.
    Dashed horizontal and vertical lines denote a $\pm10$\,\% deviation from the fitted values $A = 4.094$ and $\kappa = 0.3815$.
    The velocity profile corresponds to $Re_\tau = 10^4$ and $\delta=0.1$\,\si{\meter}.}
    \label{fig: u_primitives}
\end{figure}

Figure~\ref{fig: u_primitives} shows the joint uncertainty regions of the log-law parameters at a $95$\,\% confidence level for varying uncertainty in the primitive variables, namely (a) $U_\tau$ and $U$, and (b) $z_0$ and $\Delta_z$.
Only the statistical component of the uncertainty in the mean velocity estimates is varied in this analysis, whereas the systematic component remains unchanged.
For reference, the thick black lines correspond to the baseline case from figure~\ref{fig: u_baseline}.

Uncertainties in the friction velocity, mean velocity, and reference wall datum all significantly influence the joint uncertainty region, whereas the relative positional error $\sigma_{\Delta_z}$ has minimal impact.
A $0.2$\,\% increase in $\sigma^A_{U,i}/U_i$ more than doubles $\sigma_\kappa/\kappa$ and $\sigma_A/A$, exceeding $10$\,\% and $20$\,\%, respectively, when combined with a simultaneous increase in $\sigma_{U_\tau}/U_\tau$.
The nominal values of the log-law parameters exhibit weak sensitivity to $\sigma_{z_0}$, drifting slightly towards lower values with increasing uncertainty, but remain unchanged for variations in $\sigma_{U_\tau}/U_\tau$ and $\sigma^A_{U,i}/U_i$.
The contribution from $\sigma_{\Delta_z}$ is only marginal because, at $Re_\tau = 10^4$ and $\delta=0.1$\,\si{\meter}, it ranges between $0.1$ and $1$ viscous units, a negligible amount relative to the location and extent of the log region that spans across $300<z^+<1200$.
Its effect is expected to become more pronounced at higher $Re_\tau/\delta$.

The log-law parameters within the joint uncertainty region exhibit a strong correlation, depending on the uncertainties in the primitive variables.
Specifically, they become positively correlated with increasing  $\sigma_{U,i}/U_i$ and $\sigma_{z_0}$, producing a compounding effect, and become negatively correlated with increasing $\sigma_{U_\tau}/U_\tau$.
This behaviour arises from the functional relationship between the log-law parameters and the primitive variables in the regression model, expressed by equation \ref{eq: linear_model}.
As the partial derivatives $\partial A / \partial U$, $\partial \kappa / \partial U$, $\partial A / \partial z_0$, and $\partial \kappa / \partial z_0$ are all positive, the correlation coefficient $\rho_{A, \kappa}$, which is proportional to the product $(\partial A / \partial \theta)(\partial \kappa / \partial \theta)$ for any arbitrary variable $\theta$, becomes positive when $\sigma_{U,i}/U_i$ and/or $\sigma_{z_0}$ dominate.
Conversely, $\partial A / \partial U_\tau < 0$ and $\partial \kappa /  \partial U_\tau > 0$ yield a correlation coefficient $\rho_{A, \kappa} < 0$ when $\sigma_{U_\tau}/U_\tau$ is instead the dominant contribution.

\subsubsection{Flow scale}

\begin{figure}
    \centering
    \includegraphics{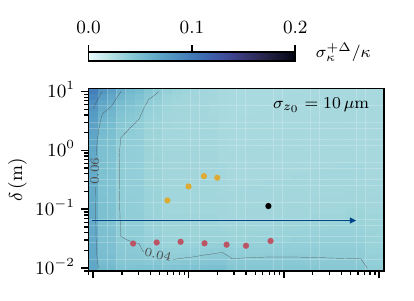}
    \includegraphics{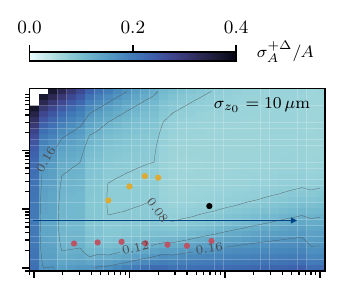}
    \includegraphics{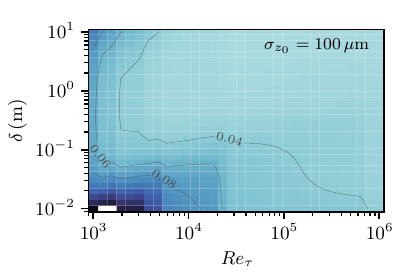}
    \includegraphics{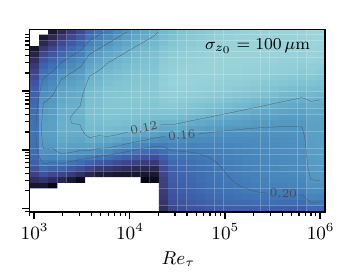}
    \caption{Maps of the positive marginal uncertainties in the log-law fit parameters $\sigma_\kappa^{+\Delta}/\kappa$ and $\sigma_A^{+\Delta}/A$, as defined in figure \ref{fig: u_baseline}b.
    The maps are presented as functions of the friction Reynolds number $Re_\tau$ and the boundary-layer thickness $\delta$, along the horizontal and vertical axes, respectively.
    Panels (a) and (c) correspond to $\sigma_{\Delta_z} = 10\,\si{\micro\meter}$, while panels (b) and (d) correspond to $\sigma_{\Delta_z} = 100\,\si{\micro\meter}$.
    Test cases from studies of high Reynolds number boundary layers and pipe flows conducted at the Princeton Superpipe facility \citep[blue line,][]{Zagarola1998, McKeon2004, Hultmark2012}, the US Navy’s William B. Morgan Large Cavitation Channel \citep[black marker,][]{Winkel2012}, the Princeton High Reynolds Number Test Facility \citep[red markers,][]{Vallikivi2015}, and the Melbourne High Reynolds Number Boundary Layer Wind Tunnel \citep[yellow markers,][]{Samie2018} are included for reference.}
    \label{fig: scale_effect}
\end{figure}

The inner and outer flow scales, characterised by the friction Reynolds number $Re_\tau$ and the boundary-layer thickness $\delta$, directly influence the propagation of uncertainty into the log-law parameters.
As evidenced in section~\ref{subsec: decomposition}, the statistical component of the covariance matrix $\bm\Sigma_\psi$, which is a weighted analogue of its \gls{ols} counterpart, scales approximately as $\delta^{2} / (Re_\tau^{2} N\sigma^2_\zeta)$.
It increases with $\delta$ but decreases with $Re_\tau$, both directly as $Re_\tau^{-2}$, and indirectly through the consequent increase in the number of data points within the log region $N$ and their spatial distribution $\sigma^2_\zeta$.
The inner and outer flow scales additionally control the significance of the systematic uncertainty arising from the reference wall datum $z_0$ and the relative positional error $\Delta z$.
Increasing $Re_\tau$ at fixed $\delta$ reduces the viscous length scale $\nu/U_\tau$, thereby increasing the viscous-normalised uncertainty contributions $\sigma^2_{z_0^+}$ and $\sigma^2_{\Delta z^+}$.
The reverse holds for increases in $\delta$ at fixed $Re_\tau$.

To illustrate the combined effect of these competing mechanisms, figure~\ref{fig: scale_effect} maps the positive marginal uncertainties $\sigma_A^{+\Delta}/A$ and $\sigma_\kappa^{+\Delta}/\kappa$, defined in figure~\ref{fig: u_baseline}, as functions of $Re_\tau$ along the horizontal axis and $\delta$ along the vertical axis, for two levels of $\sigma_{z_0}$.
The effect is more clearly visible for parameter $A$, which exhibits a markedly higher sensitivity than $\kappa$.
For $\sigma_{z_0} = 10$\,\si{\micro\meter}, the uncertainty margins in both parameters are relatively small across $Re_\tau > 1\times10^4$ and $\delta > 0.1$\,\si{\meter}, with $\sigma_\kappa^{+\Delta}/\kappa$ and $\sigma_A^{+\Delta}/A$ falling below $4$\,\% and $8$\,\%, respectively.
For $\sigma_{z_0} = 100$\,\si{\micro\meter}, they remain largely unaffected for $\delta>0.1$\,\si{\meter}, yet become exceedingly large below this threshold.
Variations in uncertainty are governed by the relative contributions of the statistical component, driven by $\delta^{2}Re_\tau^{-2} N^{-1}\sigma^{-2}_\zeta$, and the systematic component, driven by $\sigma^2_{z_0^+}$ and $\sigma^2_{\Delta z^+}$. 
The former dominates at high $\delta$ and/or low $Re_\tau$, whereas the latter dominates at low $\delta$ and/or high $Re_\tau$.
Increasing $\delta$ at fixed $Re_\tau$ first drives uncertainty down, as the systematic component becomes less significant, then back up once the statistical component, scaling as $\delta^2$ (or equivalently $U^{-2}_\tau$), takes over.
At fixed and sufficiently small $\delta$, where $\sigma^2_{z_0^+}$ and $\sigma^2_{\Delta z^+}$ remain significant, the systematic component drives uncertainty up with increasing $Re_\tau$.
This trend is eventually reduced and reversed as the statistical component, scaling as $Re_\tau^{-2} N^{-1}\sigma^{-2}_\zeta$, contributes at a comparable but opposing rate.
At large $\delta$, where $\sigma^2_{z_0^+}$ and $\sigma^2_{\Delta z^+}$ are negligible, the uncertainty decreases steeply and monotonically with $Re_\tau$, governed entirely by the statistical component.

Test conditions of benchmark experimental facilities are mapped out in figure~\ref{fig: scale_effect}, including the Princeton Superpipe facility \citep{Zagarola1998, McKeon2004, Hultmark2012}, the US Navy’s William B. Morgan Large Cavitation Channel (LCC) \citep{Winkel2012}, the Princeton High Reynolds Number Test Facility \citep{Vallikivi2015}, and the Melbourne High Reynolds Number Boundary Layer Wind Tunnel \citep{Samie2018}.
Mapping these reference cases shows that the capacity of existing facilities to achieve increasingly high Reynolds numbers yields diminishing returns in uncertainty reduction without a corresponding increase in boundary-layer thickness or better estimates of the wall-normal coordinate.
It also highlights the importance of an accurate measurement of the reference wall datum, particularly for small $\delta$, such as in \citet{Vallikivi2015}.
In this study, $\sigma_{z_0} = 5$\,\si{\micro\meter} and $\sigma_{\Delta_z} = 0.5$\,\si{\micro\meter}, half the baseline uncertainty values.

\subsubsection{Degrees of freedom}

\begin{figure}
    \centering
    \includegraphics{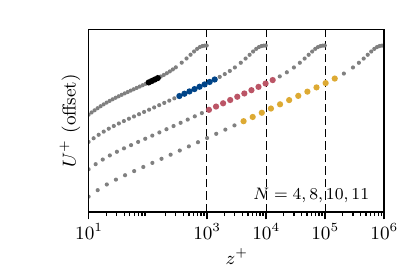}
    \includegraphics{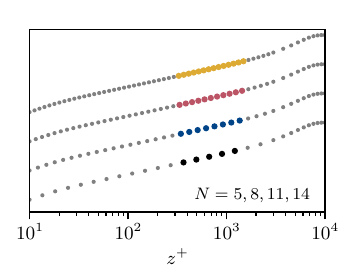}
    \caption{Synthetic hot-wire measurements of the viscous-normalised velocity profile with parameters listed in table~\ref{tab: parameters}. 
    Coloured markers denote data points within the log region $3Re_\tau^{1/2} < z^+ < 0.15Re_\tau$.
    (a) Profiles with varying friction Reynolds number $Re_\tau$, ranging from $10^3$ to $10^6$.
    Each profile comprises $30$ logarithmically spaced data points in the range $10 < z^+ < 0.3Re_\tau$, followed by $10$ linearly spaced points extending to $z^+ = \delta^+$.
    (b) Profiles at constant $Re_\tau$ with varying numbers of logarithmically spaced data points, ranging from $20$ to $50$ in increments of $10$.}
    \label{fig: resolution_bl}
\end{figure}

\begin{figure}
    \centering
    \includegraphics{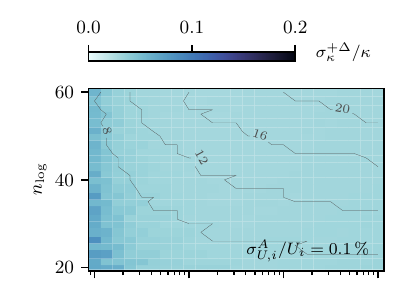}
    \includegraphics{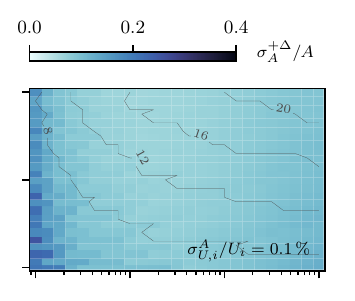}
    \includegraphics{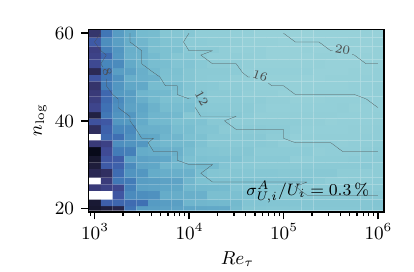}
    \includegraphics{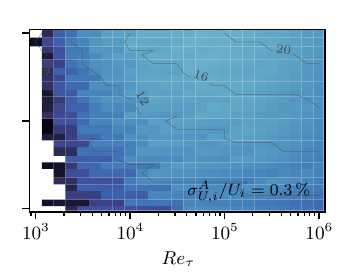}
    \caption{Maps of the positive marginal uncertainties in the log-law fit parameters $\sigma_\kappa^{+\Delta}/\kappa$ and $\sigma_A^{+\Delta}/A$, as defined in figure \ref{fig: u_baseline}b, overlaid with contours of the number of data points in the log regions $N$.
    The maps are presented as functions of the friction Reynolds number $Re_\tau$ and the number of logarithmically-spaced data points along the velocity profile $n_{\log}$, represented along the horizontal and vertical axes, respectively.}
    \label{fig: resolution_effect}
\end{figure}

The number of degrees of freedom of the regression model $N - q$ is the determined by the number of data points within the log region $N$ and the number of regression coefficients $q = 2$.
Although $N - q$ does not appear explicitly in the \gls{gls} formulation, the statistical component of the covariance matrix $\bm{\Sigma}_\psi$ scales as $\delta^{2} / (Re_\tau^{2} N\sigma^2_\zeta)$, analogously to the \gls{ols} formulation detailed in section \ref{subsec: decomposition}.
A higher $N$ can be achieved either by increasing $Re_\tau$, thereby extending the log region, or by increasing the measurement spatial resolution, as illustrated in figure~\ref{fig: resolution_bl}.
For a fixed number of data points along the velocity profile, $n_{\text{log}}=30$ logarithmically-spaced points in the range $10<z^+<0.3Re_\tau$ followed by $n_{\text{lin}}=10$ linearly-spaced points to $z^+=\delta^+$, $N$ increases steadily with $Re_\tau$, albeit with diminishing returns at higher values.
Similarly, for a fixed $Re_\tau$, increasing $n_{\text{log}}$ from $20$ to $50$ in increments of $10$ yields a comparable increase in $N$.

Figure~\ref{fig: resolution_effect} presents maps of $\sigma_A^{+\Delta}/A$ and $\sigma_\kappa^{+\Delta}/\kappa$, overlaid with contours of $N$, as functions of $Re_\tau$ and $n_{\log}$, for two values of $\sigma^A_{U,i}/U_i$, representing the primary source of statistical uncertainty.
For $\sigma^A_{U,i}/U_i = 0.1$\,\%, the uncertainty margins in the log-law parameters decrease rapidly and approach the level set by the systematic uncertainty component, with no appreciable improvement beyond $N=8$.
For $\sigma^A_{U,i}/U_i = 0.3$\,\%, the uncertainty margins are considerably higher, yet a threshold of $N=12$ still ensures a negligible contribution of the statistical component.
A steeper gradient is expected for variations in $Re_\tau$, because it acts both directly and indirectly, through the consequent growth in $N$ and the spread of data points $\sigma_\zeta^2$.

\subsubsection{Choice of log-law region}
\label{subsubsec: choice_loglaw}

Although the choice of the log region is beyond the scope of the \gls{gls} formulation, it remains an integral part of the curve-fitting procedure, with profound implications on both the estimates and uncertainties of the log-law parameters.
The effect is twofold: First, it controls the spread and the number of data points within the log region (degrees of freedom).
Second, it influences any systematic error emerging from a potential deviation of the outermost points from the logarithmic distribution.
This deviation is particularly challenging to detect because of its weak and gradual nature, occurring at levels of the order of the measurement uncertainty \citep{Marusic2013}.
More conservative bounds set a narrower log region, reducing the systematic error at the expense of fewer degrees of freedom, and vice versa.

\cite{Orlu2010} surveyed reported values for the lower and upper bounds of the log region, $z^+_{l}$ and $z^+_{u}$, respectively, revealing a substantial variation with no clear consensus.
Many studies from this survey are compiled in table~\ref{tab: survey}, along with more recent investigations.
The bounds depend on the flow conditions.
For \gls{zpg} boundary layers, the location of the upper bound is consistent across most studies and is typically assumed to lie at $z^+_{u} = 0.15 Re_\tau$.
In contrast, for channel and pipe flows, a wide range has been proposed, spanning from $0.07 Re_\tau$ \citep{Zagarola1998} up to $0.45 Re_\tau$ \citep{Jimenez2007, Monkewitz2023}.
The lower bound shows a relatively greater scatter, with $z^+_{l}$ ranging from $50$ to $400$ for boundary layers \citep{Coles1954, Nagib2008}, $100$ to $350$ for channel flows \citep{Zanoun2003, Nagib2008}, and $100$ to $600$ for pipe flows \citep{Perry2001, McKeon2004}.
Departing from the classical assumption of a fixed value, \citet{Marusic2013} also proposed the Reynolds-number-dependent lower bound $z^+_{l}=3Re_\tau^{1/2}$.
This definition stems from analytical arguments considering the balance of terms in the mean momentum equation, suggesting that the mean viscous force loses leading-order influence for $z^+ \geq 2.6Re_\tau^{1/2}$ \citep{Wei2005, Eyink2008, Klewicki2009}.

\begin{figure}
    \centering
    \includegraphics{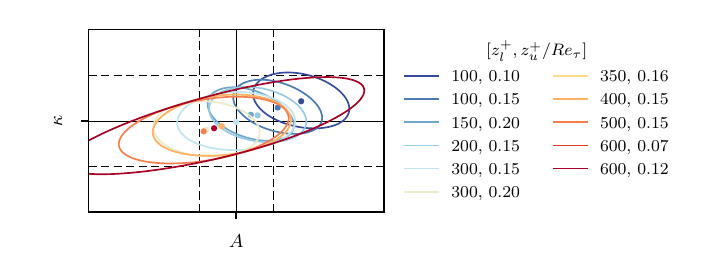}
    \caption{Estimates of the log-law parameters and the corresponding joint-uncertainty regions at a confidence level of $95$\,\% for the bounds of the log region listed in table~\ref{tab: survey}.
    The dashed vertical and horizontal lines denote a $\pm 5$\,\% deviation from the nominal values. The velocity profile corresponds to $Re_\tau = 10^4$ and $\delta=0.1$\,\si{\meter}.}
    \label{fig: log_region_survey}
\end{figure}

\begin{figure}
    \centering
    \includegraphics{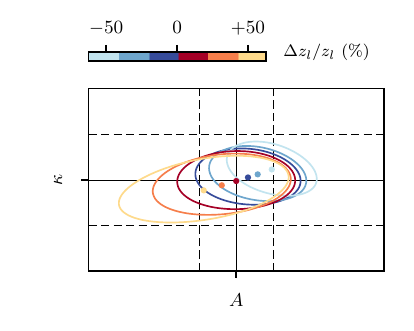}
    \includegraphics{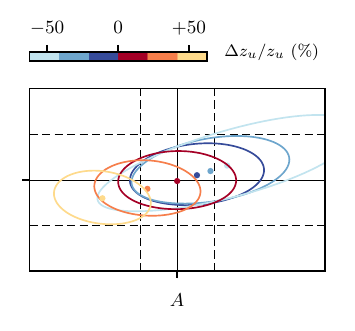}
    \caption{Estimates of the log-law parameters and the corresponding joint-uncertainty regions at a confidence level of $95$\,\%  for systematic variations of the lower and upper bounds of the log region, $z^+_{l}$ (a) and $z^+_{u}$ (b), respectively.
    The reference range of the log region is $[z^+_{l}, z^+_{u}] = [3 Re_\tau ^ {1/2}, 0.15 Re_\tau]$ \citep{Marusic2013}.
    The dashed vertical and horizontal lines denote a $\pm 5$\,\% deviation from the nominal values.
    The velocity profile corresponds to $Re_\tau = 10^4$ and $\delta=0.1$\,\si{\meter}.}
    \label{fig: log_region_effect}
\end{figure}

Figure~\ref{fig: log_region_survey} shows the estimates of the log-law parameters and joint-uncertainty regions for the reported bounds listed in table~\ref{tab: survey}.
It excludes $[z^+_{l}, z^+_{u}]=[300, 0.45 Re_\tau]$ \citep{Jimenez2007} and $[1200, 0.45 Re_\tau]$ \citep{Monkewitz2023}, which are more suitable for channel or pipe flows.
These flows are characterised by a less pronounced wake region than the composite velocity profile formulation adopted in this analysis \citep[see section \ref{subsec: gen_data} and ][]{Nagib2008}.
The estimates of the log-law parameters exhibit considerable scatter, primarily from the influence of the outermost points in the regression, which tend to deviate from a logarithmic distribution.
The corresponding joint uncertainty regions also vary significantly, depending on the relative contributions of the statistical and systematic uncertainty components.
For conservative bounds, such as $[600, 0.07 Re_\tau]$, both the spread and the number of data points in the log region are restricted.
As a result, the statistical component driven by $\bm\Sigma^A_{U}$ dominates and $\rho_{\kappa A}>0$, as discussed in section~\ref{subsubsec: u_primitives}.
Conversely, for relaxed bounds, such as $[100,\,0.15\,Re_\tau]$, the systematic component driven by $\sigma^2_{U_\tau}$ becomes more significant, yielding $\rho_{\kappa A}<0$.

A more systematic assessment of the sensitivity of the log-law parameters and joint uncertainty regions is presented in figure~\ref{fig: log_region_effect}, for deviations up to $\pm50$\,\% from the reference bounds $[z^+_{l}, z^+_{u}] = [3 Re_\tau^{1/2}, 0.15 Re_\tau]$, which correspond to $[300,1500]$ at $Re_\tau = 10^4$.
The estimates of the log-law parameters drift gradually for both positive and negative deviations of the lower and upper bounds, with $A$ exhibiting considerably greater relative sensitivity than $\kappa$.
For example, a $50$\,\% deviation in $z^+_{l}$ causes $A$ to drift by approximately $5$\,\%, while $\kappa$ drifts by less than $2$\,\%.
A similar trend is observed for deviations in $z^+_{u}$.
Deviations that extend the log region yield smaller uncertainty regions with negative correlation ($\rho_{\kappa A}<0$), while deviations that narrow the log region yield larger uncertainty regions with positive correlation ($\rho_{\kappa A}>0$).
\section{Insights and practical considerations}
\label{sec: insights}

Having defined the \gls{gls} regression model of the log law and examined its response to various sources of uncertainty, section~\ref{subsec: kappaA-kappa} extends the analysis to the correlation between the fit parameters and explores its connection to the empirical $A$--$\kappa A$ relationship introduced by \citet{Nagib2008}.
The following section~\ref{subsec: best_uncertainty} attempts to establish best estimates of the log-law parameters under the most favourable uncertainty budget, introducing an objective approach for prescribing the bounds of the log region.

\subsection{\texorpdfstring{$A$--$\kappa A$ relationship}{}}
\label{subsec: kappaA-kappa}

\cite{Chauhan2007} and \cite{Nagib2008}, following a comprehensive analysis of high-Reynolds-number wall-bounded flows, questioned the universal nature of the log-law parameters. Their analysis examined boundary-layer flow data under varying pressure-gradient conditions from the \gls{ndf} at Illinois Tech and pipe flow data from the Princeton Superpipe facility.
The log-law parameters were evaluated by fitting a composite formulation of the vertical profile of the mean streamwise velocity, valid from the wall to the edge of the boundary layer or the flow symmetry plane/axis.
Estimates of $\kappa$ appeared to vary both with pressure gradient and flow geometry, converging to measurably distinct values (within uncertainty) as the Reynolds number increased: $0.384$ for \gls{zpg} boundary-layer flows, $0.37$ for channel flows, and $0.41$ for pipe flows, with corresponding variations in $A$. 
More generally, for boundary-layer flows, estimates of $\kappa$ were noticeably larger under favourable than under adverse pressure gradients.

The view that the von Kármán constant is not universal has received support from subsequent studies such as \cite{Dixit2008} and \cite{Bourassa2009}, yet remains a contentious topic, with notable works arguing for universality.
Amongst them, \citet{Marusic2013} presented a reanalysis of several high-Reynolds-number datasets of boundary-layer, channel, and pipe flows, revealing a nominal $\kappa=0.39$ within experimental uncertainty.
Following a theoretical description of the mean flow, \cite{Klewicki2013} and \cite{Klewicki2014} demonstrated that an asymptotic value of $\kappa$ is associated with an emerging condition of dynamic self-similarity on an interior inertial domain, characterised by a geometrically self-similar hierarchy of scaling layers.
Their analysis suggested that $\kappa \equiv \phi^{-2} = 0.381\,966...$, where $\phi = (3 - 5^{1/2})/2$ is the golden ratio, would be universal.
This value is remarkably close to $\kappa = 0.372$, obtained decades earlier by \cite{Yakhot1986} using renormalisation-group methods applied to the Navier–Stokes equations.
Additionally, \cite{Luchini2017} proposed that the dependence on flow geometry at moderate Reynolds numbers observed by \cite{Nagib2008} is an effect of the pressure gradient.
They modelled the effect through a higher-order perturbation term $\beta^* (z^+/Re_\tau)$\footnote{The asterisk (*) indicates a variable from the cited reference, which is already assigned a different meaning in this manuscript.}, where $\beta^* = -(L/\tau_w)(dp/dx) = 4L/D_H$ is a geometric parameter that emerges naturally from dimensional arguments ($dp/dx$ is the streamwise pressure gradient, $\tau_w$ is the wall shear stress, $L$ is the wall-normal distance to the flow symmetry plane/axis, and $D_H$ is the hydraulic diameter).
According to their analysis, $\kappa = 0.392$ is asymptotically universal across different flow geometries, with deviations accounted entirely by the effect of the pressure gradient, which becomes negligible at higher Reynolds numbers.

These empirical and theoretical studies have since been challenged by \citet{Monkewitz2017, Monkewitz2021} and \citet{Monkewitz2023}.
Motivated by observations that a pure log distribution emerges only at extreme Reynolds numbers \citep{Zagarola1998}, they proposed a double log-layer structure for parallel flows based on matched asymptotic expansions. 
They argued that, at the moderate Reynolds numbers typical of laboratory experiments, the mean velocity profile, from which the log-law parameters are estimated, exhibits a prominent ``hump''or ``bulge'' for $z^+ \lesssim 1000$ that belongs to the inner expansion rather than to a true overlap region.
The true overlap region emerges only for $z^+ \gtrsim 1200$, extending to $z/L \sim 0.4$--$0.5$, where the mean velocity profile comprises both a log law and a linear term $S^*_0 (z^+/\mathrm{Re}_\tau)$.
In contrast to \cite{Luchini2017}, who prescribes a functional form for $S^*_0$, they estimate this parameter along with $\kappa$ by fitting velocity data from several benchmark experimental studies.
The fit yields $\kappa=0.384$ with negligible linear contribution for \gls{zpg} boundary-layer flows, $\kappa = 0.417$ and $S^*_0 = 1.15$ for channel flows, $\kappa = 0.433$ and $S^*_0 = 2.5$ for pipe flows.
From these results, they conclude that, although $S^*_0$ is roughly proportional to the pressure gradient parameter $\beta^*$, consistent with \cite{Luchini2017}, $\kappa$ is not universal but depends on flow geometry.

Independent of whether the log-law parameters are truly universal, the substantial scatter in reported values, even from reanalyses of identical datasets, poses a fundamental challenge to reaching consensus.
Crucially, the scatter is not random but reflects a strong correlation between $\kappa$ and $A$, as demonstrated by \cite{Nagib2008}.
They plotted $A$--$\kappa A$ pairs obtained by fitting a composite formulation of the mean velocity profile with a simple log overlap to various measurements of wall-bounded flows, as described above, revealing that the data distribution closely conforms to the empirical relationship
\begin{equation}
    \kappa A = 1.6 \,[\exp(0.1663A) - 1].
    \label{eq: A-kA_Nagib}
\end{equation}
The dataset included estimates of non-\gls{zpg} boundary-layer flows from \cite{Chauhan2007}, with favourable pressure gradients systematically producing higher $\kappa$ and $A$ values than adverse pressure gradients.
According to \cite{Nagib2008}, equation~\ref{eq: A-kA_Nagib} provides visual evidence for the non-universality of $\kappa$. 
A truly universal $\kappa$ would presumably produce a linear relationship, whereas the exponential behaviour suggests otherwise.
This equation was recently revisited by \cite{Baxerres2024}, who re-evaluated the log-law parameters using instead the log-linear overlap formulation derived by \citet{Monkewitz2023}, yielding a similar expression with different coefficients,
\begin{equation}
    \kappa A = 4.05 \,[\exp(0.087A) - 1].
    \label{eq: A-kA_Baxerres}
\end{equation}
Alternatively, \cite{Zanoun2026} recently proposed a linear relationship, expressed as
\begin{equation}
    \kappa A = 0.4582 A - 0.3134.
    \label{eq: A-kA_Zanoun}
\end{equation}
Following a similar argument to that of \citet{Luchini2017}, they established that $\kappa=1/e\approx0.367\,879...$ is asymptotically universal across different wall-bounded flow geometries when the pressure gradient effect is properly accounted for.
Without an explicit pressure gradient correction term, however, the log-law parameters fall along the linear trend described by equation~\ref{eq: A-kA_Zanoun}.

\begin{figure}
    \centering
    \includegraphics{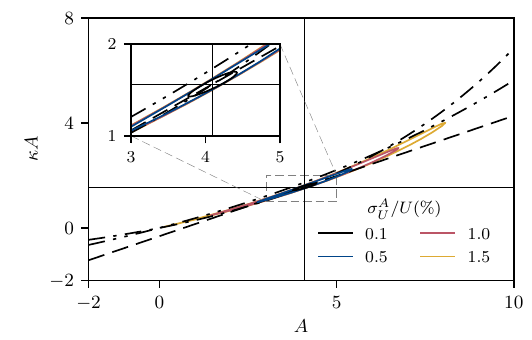}
    \caption{$A$--$\kappa A$ relationships from \citet{Nagib2008} (equation~\ref{eq: A-kA_Nagib}, $-\cdot$), \citet{Baxerres2024} (equation~\ref{eq: A-kA_Baxerres}, $-\cdot\cdot$), and \citet{Zanoun2026} (equation~\ref{eq: A-kA_Zanoun}, $--$), overlaid with joint uncertainty regions of the log-law parameters at $95$\,\% confidence level for the baseline uncertainty values in the primitive variables and $\sigma^A_U/U$ ranging from $0.1$ to $1.5$\,\%.
    The velocity profile corresponds to $Re_\tau = 10^4$ and $\delta=0.1$\,\si{\meter}.}
    \label{fig: kA-A}
\end{figure}

Figure~\ref{fig: kA-A} compares the empirical $A$--$\kappa A$ relationships from equations~\ref{eq: A-kA_Nagib}--\ref{eq: A-kA_Zanoun} against the joint-uncertainty regions of the log-law parameters for the baseline uncertainty values in the primitive variables and $\sigma^A_{U,i}/U_i$ ranging from $0.1$ to $1.5$\,\%.
The joint uncertainty regions exhibit a striking similarity to the revised relationship proposed by \citet{Baxerres2024}.
This similarity suggests that part of the scatter in reported log-law parameters likely arises from measurement uncertainty.
Notably, the spread of the data reported by \citet{Nagib2008} and \citet{Baxerres2024} for pipe and channel flows, as well as boundary-layer flows under mild pressure gradients (not included in figure~\ref{fig: kA-A}), falls approximately within the uncertainty margins for $\sigma^A_{U,i}/U_i = 0.5$\,\%.
Still, some data from non-\gls{zpg} boundary layers, presumably corresponding to extreme cases, scatter beyond the most conservative uncertainty margins, and could only be explained by an unreasonably large $\sigma^A_{U,i}/U_i$.
The present analysis, therefore, does not rule out the possibility of systematic variations from an effect of pressure gradient at finite Reynolds number.

\begin{figure}
    \centering
    \includegraphics{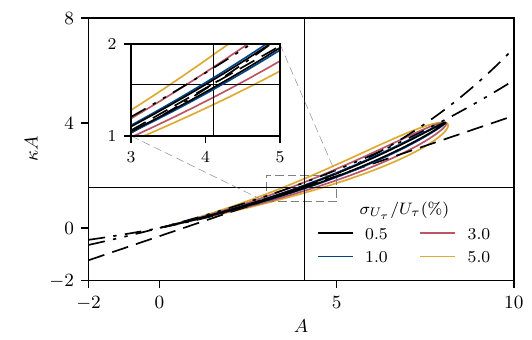}
    \caption{$A$--$\kappa A$ relationships from \citet{Nagib2008} (equation~\ref{eq: A-kA_Nagib}, $-\cdot$), \citet{Baxerres2024} (equation~\ref{eq: A-kA_Baxerres}, $-\cdot\cdot$), and \citet{Zanoun2026} (equation~\ref{eq: A-kA_Zanoun}, $--$), overlaid with joint uncertainty regions of the log-law parameters at $95$\,\% confidence level for the baseline uncertainty values in the primitive variables, $\sigma^A_U/U = 1$\,\%, and $\sigma_{U_\tau}/U_{\tau}$ ranging from $0.5$ to $5$\,\%.
    The velocity profile corresponds to $Re_\tau = 10^4$ and $\delta=0.1$\,\si{\meter}.}
    \label{fig: kA-A_U_tau}
\end{figure}

The empirical $A$--$\kappa A$ relationship emerges naturally from the uncertainty analysis, because the regression model admits families of solutions that vary analogously to the velocity profile in the log region under non-\gls{zpg} conditions.
The joint uncertainty region is elongated along the direction least constrained by the regression, specifically, the direction along which the log-law parameters co-vary with minimal change to the residuals.
In physical terms, this variation corresponds to a gradual, continuous change in shear, which closely mirrors the effect of the pressure gradient reported by \citet{Nagib2007} and \citet{Baxerres2024}.
Incidentally, the correspondence between the empirical relationships and the joint uncertainty region makes it difficult to ascertain whether the underlying trend is exponential or strictly linear.
This ambiguity is better illustrated in figure \ref{fig: kA-A_U_tau}, which plots the joint uncertainty regions for the baseline uncertainty values in the primitive variables, $\sigma^A_{U,i}/U_i= 0.5\,$\%, and $\sigma_{U_\tau}/U_{\tau}$ ranging from $1$ to $5$\,\%.
The most extreme value of $\sigma_{U_\tau}/U_{\tau}$ is representative of the uncertainty associated with an indirect assessment of the friction velocity.

In section \ref{subsec: decomposition}, decomposing the \gls{ols} covariance matrix into geometry- and flow- dependent terms reveals that the correlation between the log-law parameters depends on the Reynolds number.
Assuming the statistical component is the primary contributor to the uncertainty budget, it is then possible to derive an analytical expression for a $A$--$\kappa A$ relationship corresponding to the major axis of the joint-uncertainty region.
Approximating equation \ref{eq: Sigma_phi} for moderate $Re_\tau$, such that $R>\overline{\zeta}$ but not asymptotically dominant, the ratio $\overline{S}_2/(NR^2)=\mathcal{O}(\overline{\zeta}^2/R^2)$ is negligible, whereas $\overline{S}_1/R=\mathcal{O}(N\overline{\zeta}/R)$ remains small but finite. 
Retaining terms up to first order in $\overline{S}_1/R$, the covariance matrix reduces to
\begin{align}
    \bm{\Sigma}_\psi = \frac{\sigma^2_U}{U^2_\tau N^2\sigma^2_{\zeta}} \begin{bmatrix} 2R\overline{S}_1 + NR^2 & NR\kappa^2 \\ NR\kappa^2 & \kappa^4 N \end{bmatrix}.
\end{align}
The moderate-$Re_\tau$ approximation is considered here, rather than the asymptotic high-$Re_\tau$ limit, to achieve better accuracy while preserving analytical tractability.
According to this formulation, the principal (major-axis) direction satisfies
\begin{equation}
    \frac{d\kappa}{dA} = \frac{\kappa^2}{R}\left(1 - \dfrac{\overline{S}_1}{NR}\right).
\end{equation}
Integrating and rearranging yields
\begin{equation}
    \kappa A = -\,\frac{A}{\dfrac{A}{R} \left( 1 - \dfrac{\overline{S}_1}{NR} \right) + C^{te}}.
    \label{eq: A-kA_OLS}
\end{equation}
with the integration constant determined by the reference fit parameters $(A_0, \kappa_0)$,
\begin{equation}
    C^{te} = - \left[ \frac{1}{\kappa_0} + \frac{A_0}{R} \left( 1 - \frac{\overline{S}_1}{NR}\right) \right].
\end{equation}

\begin{figure}
    \centering
    \includegraphics{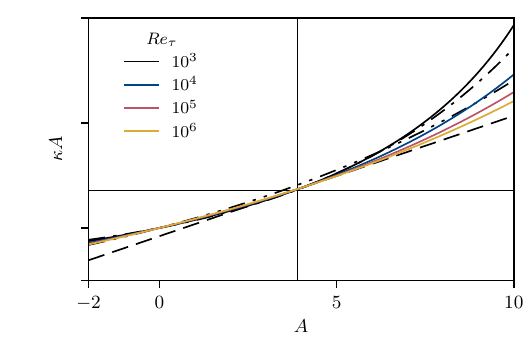}
    \caption{$A$--$\kappa A$ relationships from \citet{Nagib2008} (equation~\ref{eq: A-kA_Nagib}, $-\cdot$), \citet{Baxerres2024} (equation~\ref{eq: A-kA_Baxerres}, $-\cdot\cdot$), and \citet{Zanoun2026} (equation~\ref{eq: A-kA_Zanoun}, $--$), overlaid with the major axis of the joint-uncertainty regions for varying $Re_\tau$, as expressed by equation \ref{eq: A-kA_OLS}.
    The velocity profile corresponds to $\delta=0.1$\,\si{\meter}.}
    \label{fig: kA-A_Re_tau}
\end{figure}

Equation \ref{eq: A-kA_OLS} is plotted in figure \ref{fig: kA-A_Re_tau} for varying $Re_\tau$, alongside the empirical $A$--$\kappa A$ relationships.
For high-$Re_\tau$, it reduces to
\begin{equation}
    \kappa A = \frac{A}{\dfrac{1}{\kappa_0} - \dfrac{A - A_0}{R}},
    \label{eq: A-kA_OLS_high_Re_tau}
\end{equation}
approaching a linear trend as $R^{-1} \rightarrow 0$.
Remarkably, equation~\ref{eq: A-kA_OLS} conforms to the empirical $A$--$\kappa A$ relationships across a range of $Re_\tau$, collapsing with equation~\ref{eq: A-kA_Nagib} \citep{Nagib2008} at $Re_\tau \sim 10^3$ and equation~\ref{eq: A-kA_Baxerres} \citep{Baxerres2024} at $Re_\tau \sim 10^4$.
This result implies that severe scatter in the $A$--$\kappa A$ space arising from measurement uncertainty will tend to exhibit exponential-like behaviour at low Reynolds numbers and linear behaviour at high Reynolds numbers.
Yet, for narrow uncertainty margins, these trends are hardly distinguishable, and the scatter remains approximately linear.

It is worth noting that there is no clear reason why the reported empirical relationships should correspond so closely to the statistical correlation between the log-law parameters.
The former are presumably a consequence of the physical influence of the pressure gradient at finite Reynolds numbers, whereas the latter is a purely statistical result.
Despite their fundamental difference, the agreement is striking, and it is tempting to suggest that the functional form of equation~\ref{eq: A-kA_OLS_high_Re_tau} could serve as an alternative to the proposed exponential expressions, requiring only a single fitting parameter in place of $R$, rather than two.
A related observation is that researchers have proposed using the empirical relationships to reduce the two-point fit of the velocity profile to a one-point fit.
However, this approach implicitly assumes that the empirical relationships and the statistical correlation between the log-law parameters are the same curve.
They are coincidentally close for narrow uncertainty margins, but a perfect correspondence should not be expected, especially considering the dependence of the statistical correlation on $Re_\tau$.
Additionally, it assumes that all $A$--$\kappa A$ pairs along the curve are equally valid solutions to the linear regression, which is incorrect, as they conform to a normal probability distribution with a single most likely solution.
Fixing $A$ to estimate $\kappa$ does not improve the uncertainty margins.
It simply reduces the two-dimensional joint uncertainty region to a one-dimensional slice at a prescribed value of $A$.

\subsection{As certain as uncertainty gets}
\label{subsec: best_uncertainty}

How close can estimates of the log-law parameters get?
To address this question, the present analysis uses the lowest reported uncertainties for the primitive variables discussed in section~\ref{subsec: baseline}, reflecting the current limitations of experimental methods.
These values largely coincide with the reasonably optimistic baseline values listed in table~\ref{tab: uncertainty}, most of which remain unchanged, except for $\sigma_{z_0} = 5$\,\si{\micro\meter}, $\sigma_{\rho} = 0.003\,\rho$, $\sigma_{\nu} = 0.006\,\nu$, and $\sigma_{U_\tau} = 0.005\,U_\tau$.
Beyond uncertainty in the primitive variables, the bounds of the log-law region have traditionally been prescribed empirically, introducing a systematic error that cannot be readily quantified, as demonstrated in section \ref{subsubsec: choice_loglaw}.
Recognising that even the most rigorous uncertainty propagation is incomplete without accounting for this contribution, the present analysis determines the bounds by minimising the uncertainty in the estimated log-law parameters subject to a statistical consistency constraint on the fit residuals.

In the \gls{gls} framework, the uncertainty in the log-law parameters is estimated from the propagated covariance matrix $\bm{\Sigma}_{\varepsilon}$, rather than inferred from the fit residuals.
This allows assessment of the statistical admissibility of any candidate fitting window based on the $p$-value of the chi-square merit function.
The $p$-value is defined as the probability of observing a $\chi^2$ statistic as large or larger than the one computed, under the null hypothesis that the residuals are drawn from the distribution prescribed by $\bm{\Sigma}_{\varepsilon}$. 
A value near $0.5$ indicates that the residuals are fully consistent with the uncertainty model.
Values below $0.5$ indicate that the residuals exceed the model predictions, suggesting that measurement uncertainties are underestimated, whereas values above $0.5$ indicate that the residuals fall short of predictions, and the uncertainty budget is likely overestimated.

Following this approach, the optimal fitting region is defined as the solution to the minimisation problem
\begin{equation}
    \left(i_\ell^*, i_h^*\right) = \underset{(i_\ell, i_h) \,\in\, \Omega}{\arg\min} 
    \; \mathcal{C}(i_\ell, i_h),
    \label{eq: objfun}
\end{equation}
where the cost function $\mathcal{C}: \Omega \rightarrow \mathbb{R}$
\begin{equation}
    \mathcal{C}(i_\ell, i_h) =  \det\left(\mathbf{\bm\Sigma_\psi}\right) N + \mathcal{P}(p).
    \label{eq: cost}
\end{equation}
The first term in equation \ref{eq: cost}, $\det(\bm{\Sigma}_\psi) = \sigma_\kappa \sigma_A (1 - \rho_{\kappa,A}^2)^{1/2}$ is proportional to the area of the joint uncertainty region of the log-law parameters \citep{Kiefer1959}.
It is scaled linearly with the number of data points $N$ to promote more conservative fitting windows.
This formulation stabilises the solution, favouring a slight overestimation of the statistical uncertainty component.
In particular, when the propagated covariance matrix $\bm{\Sigma}_{\varepsilon}$ overestimates the fit residuals, the $p$-value is driven towards unity, which would otherwise bias the solution towards unrealistically wide fitting bounds.
Scaling by $N$ mitigates this effect.
The second term in equation \ref{eq: cost} is the penalty function
\begin{equation}
    \mathcal{P}(p) = \begin{cases} 
        \exp\left(0.1 - p\right) & p < 0.1 \\ 
        0                              & 0.1 \leq p \leq 0.9 \\ 
        \exp\left(p - 0.9\right)  & p > 0.9,
    \end{cases}
\end{equation}
which increases exponentially when the $p$-value of the chi-square merit function falls outside the acceptance interval $[0.1,\, 0.9]$.
The $p$-value is evaluated over the index window $[i_\ell, i_h]$ within the feasible domain
\begin{equation}
    \Omega = \left\{(i_\ell, i_h) \in \mathbb{Z}^2: 
    z^+(i_\ell) \geq 100, \;
    z^+(i_h)   \leq 0.5\, Re_\tau, \;
    i_h - i_\ell \geq N_{\min}
    \right\},
\end{equation}
where $N_{\min} = 5$ is the minimum number of points required for a statistically meaningful fit.

\begin{figure}
    \centering
    \includegraphics{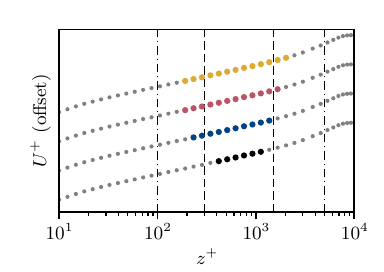}
    \includegraphics{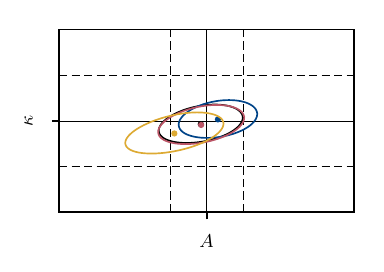}
    \caption{Solutions to the minimisation problem defined in equation~\ref{eq: objfun} as a function of the statistical uncertainty in the mean velocity $\sigma^A_{U,i}/U_i$. 
    (a) Velocity profile at friction Reynolds number $Re_\tau = 10^4$ and boundary-layer thickness $\delta = 0.1$\,\si{\meter}, with data points falling within the optimal fitting window coloured as listed in table~\ref{tab: optimisation}.
    Vertical dashed lines mark the log-law region, ranging $3Re_\tau^{1/2} < z^+ < 0.15Re_\tau$, and dash-dot lines mark the upper and lower bounds of the search domain.
    (b) Joint-uncertainty region of the log-law parameters at a $95$\,\% confidence interval.
    The solid lines denote the expected values $A = 4.17$ and $\kappa = 0.384$ (refer to table \ref{tab: parameters}), and dashed lines denote a $\pm 5$\,\% deviation thereof.}
    \label{fig: optimisation}
\end{figure}

To illustrate the application of the minimisation problem under realistic conditions, an independent Gaussian perturbation of zero mean and variance $\sigma_i^2$ is added to the noise-free composite velocity profile $U^+(z_i^+;\kappa, A, \Pi)$, defined in section~\ref{sec: synthetic},
\begin{equation}
    \tilde{U}^+(z_i^+, \sigma^2_{U,i};\kappa, A, \Pi) = 
    U^+(z_i^+;\kappa, A, \Pi) + \mathcal{N}\!\left(0,\, \sigma_i^2\right).
\end{equation}
The solutions for statistical uncertainties in the mean velocity $\sigma^A_{U,i}/U_i$ ranging from $0.05$\,\% to $0.2$\,\% are summarised in table~\ref{tab: optimisation} and figure~\ref{fig: optimisation}. 
The Gaussian perturbation remains unchanged with variance $\sigma_i/U_i^+ = 0.001$, representing the residual level expected when fitting over the optimal log region.
For $\sigma^A_{U,i}/U_i = 0.05$\,\%, the residuals are underestimated, yielding a low $p$-value and a narrow log region.
For $\sigma^A_{U,i}/U_i > 0.1$\,\%, the residuals are instead overestimated.
As a result, the optimal fitting window extends towards the limits of the feasible domain and, in extreme cases, into regions where the velocity profile visibly departs from a logarithmic distribution.
Provided these limits (indicated by the vertical dashed-dotted lines) are not reached, the residuals remain fully consistent with the assumed uncertainty model, with the $p$-value within the acceptance band, ideally close to $0.5$.
This behaviour highlights the importance of a correct estimation of the statistical uncertainty component when applying this curve-fitting procedure.
For $\sigma^A_{U,i}/U_i = 0.1$\,\%, the assumed uncertainty matches the Gaussian perturbation, yielding an optimal log region in which the residuals are consistent with the uncertainty model.

\newcommand{\circmarker}[1]{\tikz\draw[#1, fill=#1] (0,0) circle (0.6ex);}
\definecolor{tol_blue}{RGB}{0, 68, 136}
\definecolor{tol_red}{RGB}{187, 85, 102}
\definecolor{tol_yellow}{RGB}{221, 170, 51}

\begin{table}
    \centering
    \renewcommand{\arraystretch}{1.2}
    \setlength{\tabcolsep}{5mm}
    \begin{tabular}{rcccc}
        $\sigma^A_{U,i}/U_i$\,(\%)  & $0.05$        & $0.1$     &  $0.15$   & $0.2$ \\ 
        & \circmarker{black}        & \circmarker{tol_blue}     & \circmarker{tol_red} & \circmarker{tol_yellow} \\\hline
        p-value                     & $0.1088$      & $0.4361$  & $0.3180$  & $0.1002$ \\
        $\kappa$                    & $0.3826$    & $0.3846$  & $0.3822$  & $0.3787$ \\
        $A$                         & $4.137$       & $4.234$   & $4.139$   & $3.987$ \\
        $\sigma^{+\Delta}_\kappa$   & $0.0083$      & $0.0081$  & $0.0084$  & $0.0089$ \\
        $\sigma^{+\Delta}_A$        & $0.237$       & $0.222$   & $0.2428$  & $0.277$ \\
        $\rho_{\kappa, A}$          & $0.3362$      & $0.2797$  & $0.3631$  & $0.4679$
    \end{tabular}
    \caption{Summary of the optimal log-law fit as a function of the statistical uncertainty in the mean velocity $\sigma^A_{U,i}/U_i$.}
    \label{tab: optimisation}
\end{table}

The estimates of the log-law parameters closest to the reference values of $\kappa = 0.3840$ and $A = 4.170$ \citep{Musker1979, Monkewitz2007, Chauhan2009} are obtained for $\sigma^A_{U,i}/U_i = 0.1$\,\%, with $\kappa = 0.3846$ and $A = 4.234$ over a log region spanning $200 < z^+ < 1500$. 
The upper bound coincides with the empirical criterion $0.15Re_\tau$, indicated by the vertical dashed line, commonly adopted for boundary-layer flows.
The associated uncertainties are also lowest in this particular case, representing the best possible margins given the current capabilities of experimental methods. 
In relative terms, $\sigma_\kappa/\kappa = 2.0$\,\% and $\sigma_A/A = 4.8$\,\%, with a cross-correlation coefficient $\rho_{\kappa,A} = 0.2797$.
These values are considerably larger than those reported in the literature listed in table \ref{tab: survey}, indicating that uncertainty estimates have, thus far, been systematically underestimated.
Consequently, reported log-law parameters are perceived as more significant than they actually are, making it difficult to establish meaningful cross-study comparisons.

In appendix~\ref{app: log-law_real}, this fitting procedure is applied to the velocity profile of \gls{zpg} boundary layers from \citet{Samie2018}, across the range of Reynolds numbers $Re_\tau = 6\,500$ to $20\,000$.
\citet{Samie2018} did not conduct direct measurements of the wall shear stress, estimating the friction velocity and boundary-layer thickness by fitting the composite velocity profile adopted in the present study in a least-squares sense, with $\kappa = 0.384$ and $A = 4.17$.
The resulting estimates of friction velocity agree to within $\pm 1$\,\% with \gls{fe} measurements from \citet{Baars2016} taken in the same facility under similar conditions.
The \gls{gls} formulation of the log-law, in combination with the optimisation problem outlined above, correctly estimates the expected values of the log-law parameters with uncertainty margins that are consistent across Reynolds numbers. 
In contrast, fitting over an empirically prescribed log region, specifically $3Re_\tau^{1/2} < z^+ < 0.15Re_\tau$, produces considerably larger scatter.


\section{Final remarks}
\label{sec: conclusion}

The present study applies the \gls{gls} principle \citep{Aitken1935, Sprent1966} to fit the log law to the velocity profile of turbulent wall-bounded flows.
\gls{gls} retains the fundamental structure of \gls{ols} and \gls{wls}, but differs in the treatment of the weighting matrix, defined as the inverse of the covariance matrix of the residuals $\bm W = \bm\Sigma_\varepsilon^{-1}$.
Accordingly, $\bm\Sigma_\varepsilon$ is estimated by propagating uncertainties through to the transformed variables $x_i:=U_i^+$ and $y_i:=\log(z_i^+)$.
These quantities are functions of primitive variables, namely, the wall-normal coordinate $z$, the mean streamwise velocity $U$, the friction velocity $U_\tau$, and the kinematic viscosity $\nu$, which, in turn, depend on secondary primitive variables.
As primitive variables may be shared across measurements of the same quantity and across different quantities, they introduce both auto- and cross-correlated contributions to the overall uncertainty budget.
The exact covariance structure of the regression model, therefore, depends not only on the functional form of the log law but also on the experimental methods employed.

The study develops a framework tailored for measurements of the velocity profile obtained using hot-wire anemometry in combination with a motorised linear traverse, a widely used technique in wall-bounded turbulence research.
It accounts for the primary sources of uncertainty, including those associated with the measurement of the wall-normal coordinate, the mean streamwise velocity, the kinematic viscosity of air, and the friction velocity.
In particular, the uncertainty in the wall-normal coordinate incorporates contributions from both the reference wall datum and the relative displacement of the traverse system.
The uncertainty in the mean velocity comprises a systematic component, associated with the calibration of the hot-wire, and a statistical component associated with measurement variability.
The framework was applied to synthetic data based on a composite formulation of the turbulent boundary-layer velocity profile, rather than experimental measurements, thereby enabling full control over the relevant parameters and a more transparent assessment of uncertainty propagation.

Assuming baseline uncertainty values for the primitive variables, consistent with those reported in the literature, a qualitative sensitivity analysis of the log-law fit parameters reveals that the dominant contributors are the uncertainty in the friction velocity $\sigma^2_{U_\tau}$ and the statistical uncertainty in the mean velocity estimate $\bm\Sigma^A_U$.
Other terms may nonetheless become dominant for a different combination of baseline values.
The analysis also elucidates the interplay between competing mechanisms associated with systematic and statistical uncertainty components, as the flow scale and the number of degrees of freedom vary.
Notably, increasing the Reynolds number beyond the capabilities of existing benchmark facilities, $Re_\tau = \mathcal{O}(10^4)$, does not appreciably reduce the uncertainty in the log-law parameters, especially without a corresponding increase in the outer length scale (to preserve the physical dimension of the inner length scale).
Beyond this threshold, the statistical uncertainty component, whose variance scales directly with the Reynolds number as $Re_\tau^{-2}$, and indirectly as $N^{-1}\sigma^{-2}_\zeta$, becomes negligible relative to the systematic component, which then dominates the overall uncertainty budget.
For fixed flow conditions, increasing the spatial resolution of the velocity profile raises the number of points within the log region, thereby reducing the systematic uncertainty component.
However, even for relatively poorly converged mean velocity estimates with $\sigma^A_{U,i}/U_i = 0.3\%$, using more than about 12 points yields only marginal benefits.
Despite the difficulty of quantifying the influence of choosing the location and extent of the log region, the analysis showed that it significantly affects both the values of the log-law parameters and the joint-uncertainty region.

Analysis of the correlation between the log-law parameters reveals a striking similarity to the empirical $A$-$\kappa A$ relationships proposed by \citet{Nagib2008} and \citet{Baxerres2024} at moderate Reynolds numbers, approaching linear behaviour at increasingly larger values.
This similarity partly explains the scatter reported in these studies, but it does not extend to the most extreme conditions, even under conservative uncertainty assumptions.
Thus, the possibility of systematic variations arising from pressure-gradient effects at finite Reynolds numbers cannot be ruled out \citep{Nagib2008, Monkewitz2017}.
The analysis further indicates that, under weak to moderate favourable or adverse pressure gradients, deviations in the $A$-$\kappa A$ space from nominal zero-pressure-gradient values tend to be linear and cannot be definitively ascribed to either pressure-gradient effects or measurement uncertainty, assuming baseline values.
This ambiguity is especially pronounced at high Reynolds numbers and when uncertainties in the friction velocity exceed $1\%$.
It remains unclear why the empirical $A$-$\kappa A$ relationships should correspond so closely to the statistical correlation.
Nevertheless, equation \ref{eq: A-kA_OLS_high_Re_tau}, derived from the \gls{ols} covariance matrix, can be used as an alternative to the exponential expressions, requiring only a single fitting parameter rather than two.

Finally, this study proposes a fitting procedure that avoids arbitrarily prescribed bounds of the log region, an essential aspect of curve fitting that influences the estimated log-law parameters and is not directly accounted for in the uncertainty budget.
The fitting procedure assesses the statistical admissibility of any candidate fitting window (log region) based on the $p$-value of the chi-square merit function $\chi^2$, while minimising the area of the joint uncertainty region of the log-law parameters scaled linearly with the number of data points $N$.
Scaling by $N$ prevents unrealistically wide fitting bounds when the propagated covariance matrix of the residuals $\bm\Sigma_\varepsilon$ overestimates the fit residuals.
Following this approach, the best possible uncertainty margins for the log-law parameters, given the current limitations of experimental methods, are $\sigma_\kappa/\kappa = 2.0$\,\% and $\sigma_A/A = 4.8$\,\%, with a cross-correlation coefficient $\rho_{\kappa,A} = 0.2797$.
These values are significantly higher than most uncertainty estimates reported in the literature, highlighting a systematic issue in turbulent boundary-layer research.

The manuscript is accompanied by an open-source Python implementation of the \gls{gls} log-law fit, available for download on GitHub at \url{https://github.com/ma2ferreira/gls_loglaw.git}.
The repository is self-contained.
It includes all necessary routines, and its use is illustrated through example scripts, along with detailed instructions in the README file and inline code documentation.
Adoption of this tool by the wider research community would promote a standardised, comprehensive framework for quantifying uncertainty in log-law parameters across datasets, enabling meaningful and direct comparisons between studies that have thus far remained elusive.
Contributions from users via pull requests are encouraged, pending maintainer review and approval.

\backsection[Funding]{This work was financially supported by the Engineering and Physical Sciences Research Council (EPSRC) through grant EP/W026090/1.}

\backsection[Declaration of interests]{The authors report no conflict of interest.}

\backsection[Data availability statement]{The data generated in this study will be openly available in the research data archive of the University of Southampton after publication. A Python implementation of the GLS log-law fit is also available on GitHub at \url{https://github.com/ma2ferreira/gls_loglaw.git}.}

\backsection[Author ORCIDs]{M. Aguiar Ferreira, \url{https://orcid.org/0000-0002-2428-0284}; B. Ganapathisubramani, \url{https://orcid.org/0000-0001-9817-0486}.}

\backsection[Author contributions]{M. Aguiar Ferreira (A, B, C, D, E) and B.~Ganapathisubramani (A, C, E, F) contributed as follows: A — conceptualisation, B — implementation, C — analysis, D — manuscript preparation, E — editing, and F — funding.}

\appendix
\counterwithin{table}{section}
\setcounter{table}{0}

\section{Uncertainty propagation}

\subsection{Traverse system}
\label{app: Sigma_zz}
Considering a linear traverse with a position encoder and no backlash, let $z_0$ be the reference datum corresponding to the initial position of the hot-wire in the near-wall region, and $\bm{\Delta z}$ the vector of step sizes.
The wall-normal coordinate vector is then expressed as
\begin{equation}
    \bm{z} = z_0\bm{1}_n + \bm{\Delta z}\,\mathbf{L},
\end{equation}
where $\mathbf{L}$ (lower-triangular matrix of ones) is the discrete cumulative-sum operator.
Two primary sources of error can be identified, associated with the measurement of the reference wall datum $\sigma_{z_0}$ and the position encoder $\sigma_{\Delta _z}$.
These error contributions are propagated into the covariance of the wall-normal coordinate via the first-order Taylor series expansion
\begin{equation}
    \bm{\Sigma_{z}} = \sigma^2_{z_0} \bm{1}_{n \times n} + \sigma^2_{\Delta_z}\bm{L}\bm{L}^\top.
\end{equation}
This expression shows that for vertical sweeps away from the wall, the uncertainties in wall-normal coordinates are mutually correlated, with the covariance between any two points determined by the number of shared positioning steps from the reference wall datum --- i.e. the covariance between two arbitrary points $i$ and $j$ scales with $\min(i,j)$.

\subsection{Hot-wire velocity measurements}
\label{app: Sigma_uu}
To estimate the covariance matrix of the hot-wire velocity measurements $\bm{\Sigma}^B_{U}$, the \gls{gls} formulation outlined in section~\ref{sec: model} is applied to the King’s law calibration curve based on synthetic calibration data.
A typical calibration involves acquiring simultaneous measurements of the free-stream velocity from a hot-wire and a Pitot-static probe, mounted alongside.
The relationship between the velocity $U_0 = (2q_0/\rho)^{1/2}$, where $q_0$ is the free-stream dynamic pressure from the Pitot-static probe and $\rho$ is the air density, and the response of the hot-wire $E$ is described by the King's law formulation
\begin{equation}
    E^2 = B + C U^D,
    \label{eq: King_law}
\end{equation}
where $B$, $C$ and $D$ are empirically fit parameters.
For standard environmental conditions $d\approx1/2$, in which case, equation~\ref{eq: King_law} can be linearised by introducing the transformed variables $x_i:= E_i^2$ and $y_i:= {U_0}_i^{1/2} = (2{q_0}_i/\rho)^{1/4}$.
The corresponding linear regression model for a set of $N$ observations is then given by equation \ref{eq: linear_model} with $\bm\beta = [- B/C, 1/C]$.

Measurement errors in the hot-wire response $E$, the dynamic pressure $q$, and the air density $\rho$ are the primary sources of uncertainty and are assumed heteroscedastic and mutually independent.
These variables are collected in the vector of primitive variables $\bm{\theta} = \begin{bmatrix} v_1 & \cdots & v_N & q_1 & \cdots & q_N & \rho \end{bmatrix}
\in \mathbb{R}^{2N+1}$ with covariance matrix
\begin{equation}
    \bm{\Sigma}_\theta = 
    \begin{bmatrix}
        \text{diag}(\bm{\sigma}_V^2) & &\\
        & \text{diag}(\bm{\sigma}_{q_0}^2) \\
        & & \sigma_\rho^2 & \\
    \end{bmatrix}
    \in \mathbb{R}^{(2N+1) \times (2N+1)}.
\end{equation}
The uncertainty in the primitive variables is propagated to the transformed variables via a first-order Taylor series expansion,
\begin{equation}
    \bm{\Sigma}_\omega = \bm{J}_\omega \bm{\Sigma}_\theta \bm{J}_\omega^\top \in \mathbb{R}^{2N\times2N},
\end{equation}
where $\bm{J_\omega}$ the Jacobian matrix of $\bm{\omega}$ with respect to $\bm{\theta}$. 
The Jacobian matrix is written explicitly
\begin{equation}
    \bm{J_\omega} = \frac{\partial\bm{w}}{\partial\bm{\theta}} =
    \begin{bmatrix}
        \text{diag}(2\bm{V}) &  \bm{0}_{N\times N} & \bm{0}_{N} \\[4mm]
        \bm{0}_{N \times N} & \text{diag}\left(\dfrac{1}{4}\dfrac{\bm{x}}{\mathbf{q_0}}\right)  &  -\dfrac{1}{4}\dfrac{\bm{x}}{\rho} 
    \end{bmatrix} \in \mathbb{R}^{2N \times (2N+1)}.
\end{equation}

As the linear model is identical to that used for the log-law fit, the covariance matrix $\bm{\Sigma}_\varepsilon$ is computed analogously using equation \ref{eq: S_epsilon} and the covariance matrix of the fit parameters $\bm{\Sigma}_{\beta}$ using equation \ref{eq: S_beta}.

Finally, the covariance matrix of the hot-wire velocity measurements is computed in two steps.
First, by propagating the covariance of the fit parameters to the transformed velocity signal obtained from the calibration curve,
\begin{equation}
    \bm{\Sigma}_y = \bm{X}^\top\bm{\Sigma}_\beta\bm{X},
\end{equation}
where $\bm{X}$ is the Jacobian matrix of the response vector with respect to the regression coefficients $\bm\beta$.
This covariance matrix is then propagated to the velocity measurements,
\begin{equation}
    \bm{\Sigma}_{U}^B = \text{diag}(2\bm{y}) \, \bm{\Sigma}_y \, \text{diag}(2\bm{y}).
\end{equation}

\section{Survey of published studies}
\begin{ThreePartTable}
    \renewcommand{\arraystretch}{1.2}
    \small
    \begin{TableNotes}
        \item[a] Reanalysis of existing datasets. 
        \item[b] Based on Prandtl's mixing-length hypothesis. 
    \end{TableNotes}
    \setlength{\LTcapwidth}{\textwidth}
    \begin{longtable}{lccccccl}
        
        \caption{Summary of studies which provide estimates for the log-law parameters $\kappa$ and $A$.
        When reported, the corresponding uncertainties, $\sigma_\kappa$ and $\sigma_A$, as well as the lower and upper bounds of the log region, $z^+_{\log^-}$ and $\hat{z}_{\log^+}$, are also listed.} \\
        \label{tab: survey}
        \textsc{Author} & $\kappa$ & $\sigma_\kappa$ & $A$ & $\sigma_A$ & $z^+_{l}$ & $z^+_{u}$ & \textsc{Notes} \\ \hline
        \endfirsthead
        
        \caption[]{Summary of research studies -- continued from previous page} \\
        \textsc{Author} & $\kappa$ & $\sigma_\kappa$ & $A$ & $\sigma_A$ & $z^+_{l}$ & $z^+_{u}$ & \textsc{Notes} \\ \hline
        \endhead
        
        \hline
        \multicolumn{8}{r}{Continued on next page...} \\
        \endfoot
        
        \hline
        \insertTableNotes
        \endlastfoot

        \multicolumn{8}{l}{\textsc{Boundary-layer flow}} \\
        \cite{Karman1930} & $0.380$ & & & & & & \\
        \cite{Millikan1938} & $0.400$ & & & & & & \\
        \cite{Coles1968} & $0.410$ & & $5.00$ & & & & \\
        \cite{Osterlund1999} & $0.384$ & & $4.08$ & & $200$ & $0.15$ & \\
        \cite{Andreas2006} & $0.387$ & $0.003$ & & & & & \\
        \cite{Nagib2007} & $0.384$ & & $4.17$ & & $200$ & $0.15$ & Review\\
        \cite{Nagib2008} & $0.384$ & $0.005$ & & & $400$ & $0.15$ & Review\\
        \cite{Marusic2013}\tnote{a} & $0.390$ & $0.020$ & $4.30$ & $0.20$ & $3Re_\tau^{1/2}$ & $0.15$ & \cite{Kulandaivelu2012}\\
        --- & $0.391$ & $0.004$ & $4.34$ & $0.19$ & $3Re_\tau^{1/2}$ & $0.15$ & \cite{Hultmark2012} \\
        --- & $0.387$ & $0.004$ & $4.32$ & $0.20$ & $3Re_\tau^{1/2}$ & $0.15$ & \cite{Winkel2012} \\
        --- & $0.410$ & $0.028$ & $4.44$ & $1.83$ & $3Re_\tau^{1/2}$ & $0.15$ & \cite{Hutchins2002} \\
        
        & & & & & & & \\
        
        \multicolumn{8}{l}{\textsc{Channel flow}} \\
        \cite{Hussain1975} & $0.410$ & & $5.00$ & & & & Qualit.\\
        \cite{Moin1982} & $0.400$ & & $5.00$ & & & & LES $Re_\tau=600$; Qualit.\\
        \cite{Kim1987} & $0.400$ & & $5.50$ & & & & DNS $Re_\tau=180$; Qualit. \\
        \cite{Zanoun2003} & $0.370$ & & $3.7$ & & $150$ & $0.20$ & \\
        \cite{Monty2005} & $0.389$ & & $4.23$ & & $100$ & $0.15$ & \\
        \cite{Jimenez2007} & $0.402$ & $0.020$ & & & $300$ & $0.45$ & DNS $Re_\tau=2000$\\
        \cite{Nagib2008} & $0.370$ & & & & $400$ & $0.15$ & Review\\
        \cite{Lee2015} & $0.384$ & $0.004$ & $4.27$ & & $350$ & $0.16$ & DNS $Re_\tau=5186$\\
        
        & & & & & & & \\
        
        \multicolumn{8}{l}{\textsc{Pipe flow}} \\
        \cite{Nikuradse1932} &  $0.417$ & & $5.84$ & & & & Near wall\\
        --- &  $0.400$ & & $5.50$ & & & & Near wall and centreline\\
        --- &  $0.380$ & & & & & & $l_m=\kappa z$\tnote{b}; $Re_\tau \gtrsim 5000$ \\
        --- &  $0.400$ & & & & & & $l_m=\kappa z$; $Re_\tau \lesssim 5000$ \\
        \cite{Deissler1950} &  $0.360$ & & $3.80$ & & & & \\
        \cite{Hinze1959}\tnote{a} &  $0.361$ & & & & & & \cite{Nikuradse1932} \\
        \cite{Eggels1994} & $0.350$ & & $4.80$ & & & & DNS $Re_\tau=360$; Qualit.\\
        --- &  $0.400$ & & $5.5$ & & & & DNS $Re_\tau=360$; Qualit.\\
        \cite{Durst1995} & $0.400$ & & $5.5$ & & & & Qualit.\\
        \cite{Zagarola1998} &  $0.436$ & $0.002$ & $6.15$ & $0.12$ & $600$ & $0.07$ & \\
        \cite{Perry2001}\tnote{a} &  $0.390$ & & $4.42$ & & $100$ & $0.10$ & \cite{Zagarola1998}\\
        \cite{McKeon2004} &  $0.421$ & $0.002$ & $5.60$ & $0.08$ & $600$ & $0.12$ & \\
        \cite{Monty2005} &  $0.384$ & & $4.22$ & & $100$ & $0.15$ & \\
        \cite{Zanoun2007} &  $0.370$ & & $3.70$ & & $300$ & $0.20$ & \\
        \cite{Nagib2008} &  $0.410$ & & & & $400$ & $0.15$ & \\
        \cite{Wu2008} &  $0.362$ & & & & & & DNS $Re_\tau=1142$\\
        --- &  $0.426$ & & & & & & DNS $Re_\tau=180$\\  
        
        & & & & & & & \\
        
        \multicolumn{8}{l}{\textsc{All flow types}} \\
        \cite{Luchini2017}& $0.392$ & & $4.48$ & & $200$ & & Review \\
        \cite{Monkewitz2023} & $0.384$ & & $-$ & $-$ & $1200$ & $0.45$ & Review \\
        \cite{Zanoun2026} & $1/e$ & & $-$ & $-$ & $500$ & $0.15$ & Review \\
        \cite{Coles1954}   \\
        \cite{Kahler2005}  \\
    \end{longtable}
\end{ThreePartTable}

\section{GLS applied to published data}
\label{app: log-law_real}

This section presents the results of the \gls{gls} log-law fit to the velocity profiles of \gls{zpg} boundary layers reported by \citet{Samie2018}, across the range of Reynolds numbers $Re_\tau = 6{,}000$ to $20{,}000$.
Measurements were obtained using a $2.5$-\si{\micro\meter}-diameter hot wire, mounted on a vertical linear traverse with a step resolution of $5$\,\si{\micro\meter} and equipped with an optical encoder with a positional accuracy of $0.5$\,\si{\micro\meter}.
The reference wall datum was measured using a microscope with a resolution of $1$\,\si{\micro\meter}.
To ensure statistical convergence, each point in the velocity profile was sampled over $12\,000$--$15\,000$ boundary-layer turnover times, which, given the turbulence intensity within the log region, yields an uncertainty in the mean velocity estimates of approximately $0.1\%$.
Temperature was measured using a thermocouple connected to an OMEGA DP25 signal conditioner, with a resolution of $0.1$\,\si{\degreeCelsius} and an accuracy of $0.5$\,\si{\degreeCelsius}.
Throughout the measurements, temperature variations were within $1.5$\,\si{\degreeCelsius} at a nominal free-stream velocity of $20$\,\si{\meter\per\second}, and within $3$\,\si{\degreeCelsius} at $30$\,\si{\meter\per\second}.
The velocity profiles were corrected for temperature drift using the scheme proposed by \citet{Hultmark2010}.
The friction velocity was not measured directly but estimated by fitting the composite formulation for the velocity profile of a turbulent boundary layer of \citet{Chauhan2009}, assuming $\kappa = 0.384$ and $A=4.17$.
This approach has a reported accuracy of approximately $2$\,\%, and the resulting friction velocity estimates agree within $1\%$ with the Coles--Fernholz relation derived by \citet{Baars2016} based on direct wall shear stress measurements using a \gls{fe} balance.

\begin{figure}
    \centering
    \includegraphics{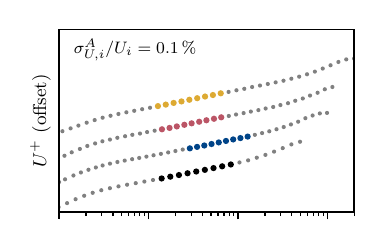}
    \includegraphics{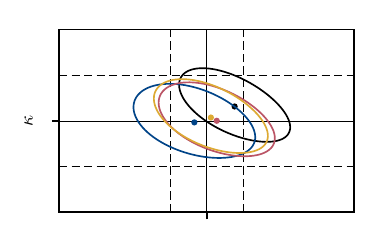}
    \includegraphics{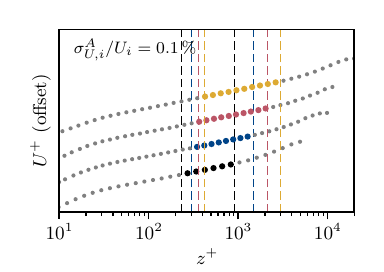}
    \includegraphics{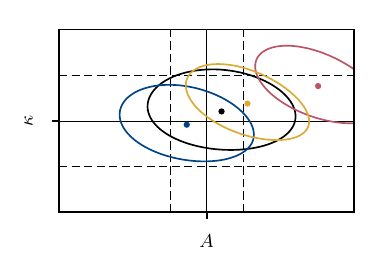}
    \caption{\gls{gls} fit of the log law to the velocity profile of \gls{zpg} boundary layers from \citet{Samie2018}, across the range of Reynolds numbers $Re_\tau = 6\,500$ to $20\,000$.
    (a) and (b) optimal bounds of the log region are determined via the minimisation problem defined in section~\ref{subsec: best_uncertainty}, whereas (c) and (d) fits over the empirically prescribed log region $3Re_\tau^{1/2} < z^+ < 0.15Re_\tau$ \citep{Marusic2013}, indicated by the coloured dashed lines.
    Joint-uncertainty regions of the log-law parameters are at a $95$\,\% confidence interval.
    The solid lines denote the expected values $A = 4.17$ and $\kappa = 0.384$, and dashed lines denote a $\pm 5$\,\% deviation thereof.
    }
    \label{fig: Samie2018}
\end{figure}

Based on these experimental details, the uncertainty in the primitive variables are as follows: $\sigma_{\Delta z} = 0.5$\,\si{\micro\meter}, $\sigma_{z_0}=2.5$\,\si{\micro\meter} is prescribed by the diameter of the hot wire, which exceeds the resolution of the microscope, $\bm\sigma^A_{U}=0.001\bm U$, $\bm\sigma_{E}=0.001\bm E$ (assumed), $\bm\sigma_{q_0}=0.001\bm q_0$ (assumed), $\sigma_{U_\tau} = 0.01 U_\tau$, $\sigma_{\rho}=0.0025\rho$, and $\sigma_{\nu}=0.005\nu$, the latter two determined by the uncertainty in temperature.
The temperature correction scheme applied to the velocity profiles represents an additional systematic uncertainty source that is not accounted in this analysis.
Figure \ref{fig: Samie2018} shows the \gls{gls} fit to the velocity profiles.
In panels (a–b), the fit is subject to the minimisation problem described in section \ref{subsec: best_uncertainty}, such that the optimal bounds of the log region ensure that residuals are consistent with the statistical component of the propagated covariance of the residuals $\bm\Sigma_\varepsilon$.
In panels (c–d), the fit is computed instead by prescribing the bounds of the log region $3Re_\tau^{1/2} < z^+ < 0.15Re_\tau$.
The optimal bounds of the log region do not follow the same trend as the prescribed bounds, with a slight shift toward the wall at the highest Reynolds numbers.
This shift likely arises from a subtle discontinuity or local variability in the velocity profile farther away from the wall, exceeding the estimated statistical uncertainty.
Although the optimal bounds may not capture the full extent of the log region, they remain within acceptable limits.
The estimates of the log-law parameters obtained for optimal bounds are consistent across Reynolds number and closely match the expected values $\kappa = 0.384$ and $A=4.17$.
In contrast, fitting the log law within the prescribed bounds produces greater scatter, which is not fully accounted for by the joint-uncertainty regions.

\bibliographystyle{jfm}
\bibliography{bibliography}

\end{document}